\documentclass[%
 aip,
 amsmath,amssymb,
 reprint,%
]{revtex4-1}

\usepackage{graphicx}
\usepackage{dcolumn}
\usepackage{bm}
\usepackage{subfigure} 
\usepackage[utf8]{inputenc}
\usepackage[T1]{fontenc}
\usepackage{mathptmx}
\usepackage{etoolbox}

\usepackage{color,soul}
\usepackage{ulem}

\makeatletter\def\@email#1#2{%
 \endgroup
 \patchcmd{\titleblock@produce}
  {\frontmatter@RRAPformat}
  {\frontmatter@RRAPformat{\produce@RRAP{*#1\href{mailto:#2}{#2}}}\frontmatter@RRAPformat}
  {}{}
}%
\makeatother
\begin{document}

\preprint{AIP/Iwamatsu2024}

\title[Flow in non-slowly-varying tube]{
A hydrodynamic model of capillary flow in an axially symmetric tube with a non-slowly-varying cross section and a boundary slip
}
\author{Masao Iwamatsu}
 \email{iwamatm@tcu.ac.jp}
\affiliation{ 
Tokyo City University, Setagaya-ku, Tokyo 158-8557, Japan
}%


\date{\today}

\begin{abstract}
The capillary flow of a Newtonian and incompressible fluid in an axially symmetric horizontal tube with a non-slowly-varying cross section and a boundary slip is considered theoretically under the assumption that the Reynolds number is small enough for the Stokes approximation to be valid.  Combining the Stokes equation with the hydrodynamic model assuming the Hagen-Poiseulle flow, a general formula for the capillary flow in a non-slowly-varying tube is derived.  Using the newly derived formula, the capillary imbibition and the time evolution of meniscus in tubes with non-uniform cross sections such as a conical tube, a power-law-shaped diverging tube, and a power-law-shaped converging tube are reconsidered.  The perturbation parameters and the corrections due to the non-slowly-varying effects are elucidated and the new scaling formulas for the time evolution of the meniscus of these specific examples are derived.  Our study could be useful for understanding various natural fluidic systems and for designing functional fluidic devices such as a diode and a switch.
\end{abstract}

\maketitle

%

\section{\label{sec:sec1}Introduction}

The capillary flow of a Newtonian and incompressible fluid~\cite{Batchelor1967} in horizontal tubes with various geometrical structure is a very general problem in fluid mechanics and appears ubiquitously in many scientific and engineering fields such as geophysics~\cite{Cai2022}, biomedical engineering~\cite{Prakash2008,Kim2012} and nano technology~\cite{Squires2005,Bocquet2010,Kavokine2021,Comanns2015}.  Even though the capillary flow~\cite{Bell1906,Lucas1918,Washburn1921} and the capillary rise~\cite{Rideal1922,Bosanquet1923} in a straight cylindrical tube have been studied for more than a century and well documented~\cite{Batchelor1967,Landau1987,deGennes2004}, the capillary flow in more complex geometry is still the issue to be investigated.

The simplest example of a non-straight tube is an axially symmetric tube with a varying cross-section radius.   The steady flow of a uniform incompressible viscous fluid in a straight cylindrical tube~\cite{Batchelor1967,Landau1987} is characterized by a parabolic velocity profile and governed by the Hagen-Poiseulle law, which was extended to the flow in a {\it slowly-varying} tube (Batchelor~\cite{Batchelor1967}, p. 217) and the volumetric flow rate $Q$ being given by
\begin{equation}
Q = \frac{\pi R(z)^{4}}{8\mu}\left(-\frac{\partial p}{\partial z}\right),
\label{eq:Sv1}
\end{equation}
where a fluid of the dynamic viscosity $\mu$ is moving in an axially symmetric tube along the $z$ axis with a
varying radius $R(z)$ under the pressure $p(z)$. The flow rate $Q$ is constant along the tube for incompressible fluids.   The tube radius $R(z)$ must be slowly varying $\left|dR/dz\right|\ll 1$; furthermore, if the condition~\cite{Batchelor1967}
\begin{equation}
\left| \frac{dR}{dz}\rho\frac{R(z)U}{\mu} \right| \ll 1,
\label{eq:Sv2}
\end{equation}
is satisfied, the inertia contribution due to the varying radius is negligible, where $\rho$ is the fluid mass density and $U$ is a representative flow speed.

Equation (\ref{eq:Sv1}) can be integrated to give
\begin{equation}
p(z_{1})-p(z_{2})=\frac{8\mu Q}{\pi}\int_{z_1}^{z_2}\frac{dz}{R(z)^{4}}.
\label{eq:Sv3}
\end{equation}
This analytical result has been rederived several times and used widely~\cite{Sharma1991,Staples2002,Young2004,Reyssat2008,Urteaga2013,Berli2014,Elizalde2014,Gorce2016,Khatoon2020,Tran-Duc2020,Iwamatsu2022} since no analytical solution of the Stokes equation for an axially symmetric tube with a varying cross-section radius is available except for a diverging conical tube~\cite{Tran-Duc2019}.

About a century ago, Washburn~\cite{Washburn1921} used the Hagen-Poiseulle law in Eq.~(\ref{eq:Sv1}) to study the capillary imbibition and the time evolution of meniscus  (flow front) in horizontal {\it straight} ($R(z)=$constant)  cylindrical tubes by assuming a quasi-steady flow and by relating the flow rate $Q$ to the meniscus velocity $dl/dt$, where $l$ being the position of meniscus.  He found a universal diffusive scaling law $l\propto t^{1/2}$, which is now known as the classical Lucas-Washburn (LW) law~\cite{Lucas1918,Washburn1921,Cai2022} or the Bell-Cameron-Lucas-Washburn (BCLW) law~\cite{Reyssat2008,Gorce2016,Bell1906,Lucas1918,Washburn1921}.  This classical LW law was extended by several authors~\cite{Reyssat2008,Urteaga2013,Berli2014,Gorce2016,Iwamatsu2022} to axially symmetric tubes with a {\it slowly} varying cross-section radius using Eqs.~(\ref{eq:Sv1}) and (\ref{eq:Sv3}) because no numerical experiments~\cite{Mondal2021} are helpful to find scaling laws.   In fact, various universal scaling laws, which are different from the LW law and depend on the tube geometry, were discovered for the capillary imbibition driven by surface tension~\cite{Reyssat2008,Urteaga2013,Berli2014,Gorce2016} and for the forced imbibition by constant external pressure~\cite{Iwamatsu2022}.  Furthermore, some of them were experimentally verified~\cite{Reyssat2008,Urteaga2013,Berli2014,Gorce2016}.

However, all those previous studies of imbibition~\cite{Sharma1991,Staples2002,Young2004,Reyssat2008,Urteaga2013,Berli2014,Elizalde2014,Gorce2016,Khatoon2020,Tran-Duc2020,Iwamatsu2022} used Eqs.~(\ref{eq:Sv1}) and (\ref{eq:Sv3})  based on the slowly-varying approximation and did not consider the non-slowly-varying effect.  How should Eq.~(\ref{eq:Sv1}) be modified when the radius $R(z)$ is not slowly varying?  What exactly is the condition other than Eq.~(\ref{eq:Sv2}) when Eq.~(\ref{eq:Sv1}) is valid?  In this paper, we consider the capillary flow in an axially symmetric tube whose cross-section radius is {\it not} necessarily slowly varying and generalize Eq.~(\ref{eq:Sv1}).  To this end, we seek for an approximate but analytical formula similar to Eq.~(\ref{eq:Sv1})  and employ the hydrodynamic model~\cite{Duarte1996,Digilov2008,Liou2009,Wang2013,Nissan2016}, which was used to derive the macroscopic equation called the variable-mass Newton equation~\cite{Bosanquet1923,Menon1987,Duarte1996,Zhmud2000,Masoodi2013} for the capillary rise from the Navier-Stokes equation.

Here, we assume that the Reynolds number is small enough for the Stokes approximation to be valid and use the same technique~\cite{Duarte1996,Digilov2008,Liou2009,Wang2013,Nissan2016} to transform the Stokes equation~\cite{Washburn1921} for the capillary flow. We adhere to the macroscopic description and include a slip length of Navier's slip boundary condition~\cite{Washburn1921,Majumder2005,Neto2005,Suk2017,Wu2017c} and neglect gravity.  We also neglect various microscopic effects near the fluid-wall interface,  such as an inhomogeneous density and viscosity, a density layering~\cite{Wu2017b,Wu2017c}, a time-dependent dynamic contact angle~\cite{Popescu2008,Wu2017}, and various effects at and near the inlet and the outlet~\cite{Sampson1891,Weissberg1962,Suk2017,Kornev2000}.

We include the slip length because it ranges from tens of nanometers to a micrometer~\cite{Majumder2005,Bocquet2010} so that it must play a crucial role in micro- and nano-fluidic systems~\cite{Mondal2021,Kavokine2021,Wu2017c}.  Furthermore, it can partly include various microscopic effects near the fluid-wall interface as an apparent slip length~\cite{Wu2017c}.  Therefore, we consider the quasi-steady capillary flow of spontaneous imbibition in a horizontal tube driven solely by a capillary force by surface tension.  We do not consider the transient effect such as the capillary rise~\cite{Rideal1922,Bosanquet1923,Popescu2008,Wu2017,Stange2003} and the droplet uptake~\cite{Willmott2011}.

This paper is organized as follows. In Sec. II, we generalize the Hagen-Poiseulle law in Eq.~(\ref{eq:Sv1}) to include the effect of a non-slowly-varying radius. Then we use this generalized formula to derive an equation of motion to describe the time evolution of the meniscus by imbibition.   In Sec. III, we use the newly derived equation of motion to study the generalization of the universal LW law in axially symmetric tubes of several typical geometries considered by previous researchers~\cite{Reyssat2008,Urteaga2013,Berli2014,Gorce2016} to understand the non-slowly-varying effect. Finally, in Sec. IV, we conclude by summarizing our new results, which will be useful for future studies of various fluidic systems made of axially symmetric tubes.

\section{\label{sec:sec2}capillary flow in an axially symmetric tube with a non-slowly-varying cross section }

Steady flow of a Newtonian and incompressible fluid with velocity $\bm{v}$ in a capillary tube is described by the Stokes equation~\cite{Batchelor1967}, which consists of the continuity equation (volume conservation law)
\begin{equation}
\bm{\nabla} \cdot \bm{v}=0,
\label{eq:Sv4}
\end{equation}
or
\begin{equation}
\frac{\partial v_{x}}{\partial x} + \frac{\partial v_{y}}{\partial y} + \frac{\partial v_{z}}{\partial z}=0,
\label{eq:Sv5}
\end{equation}
and the momentum conservation law
\begin{equation}
\mu \Delta \bm{v} = \bm{\nabla} p,
\label{eq:Sv6}
\end{equation}
or
\begin{eqnarray}
\mu\left( \frac{\partial^{2}v_{x}}{\partial x^{2}}  + \frac{\partial^{2}v_{x}}{\partial y^{2}} + \frac{\partial^{2}v_{x}}{\partial z^{2}} \right) &=& \frac{\partial p}{\partial x},
\label{eq:Sv7} \\
\mu\left( \frac{\partial^{2}v_{y}}{\partial x^{2}}  + \frac{\partial^{2}v_{y}}{\partial y^{2}} + \frac{\partial^{2}v_{y}}{\partial z^{2}} \right) &=&\frac{\partial p}{\partial y},
\label{eq:Sv8} \\
\mu\left( \frac{\partial^{2}v_{z}}{\partial x^{2}}  + \frac{\partial^{2}v_{z}}{\partial y^{2}} + \frac{\partial^{2}v_{z}}{\partial z^{2}} \right) &=& \frac{\partial p}{\partial z},
\label{eq:Sv9}
\end{eqnarray}
where we have written in the Cartesian coordinates $(x,y,z)$.  We consider a tube along the $z$ axis and a non-uniform circular cross section
\begin{equation}
x^2+y^2=R^{2}(z),
\label{eq:Sv10}
\end{equation}
or
\begin{equation}
r=R(z),
\label{eq:Sv11}
\end{equation}
with $r=\sqrt{x^2+y^2}$, where we introduced a polar coordinate system $\left(x,y\right)\rightarrow \left(r,\theta\right)$ with
\begin{equation}
x = r\cos\theta,\;\;\;y = r\sin\theta. 
\label{eq:Sv12}
\end{equation}
We assume that the capillary flow is characterized by a parabolic velocity profile~\cite{Liou2009,Wang2013,Nissan2016} with Navier's slip length $\lambda$~\cite{Neto2005,Suk2017}.  The $z$ component $v_{z}$ of the velocity at $(x,y,z)=(r,z)$ is given by~\cite{Liou2009,Wang2013,Nissan2016}
\begin{eqnarray}
v_{z}\left(z,r,t\right) &=& 2v_{l}\frac{R(l)^{2}}{R(z)^{2}}\left(1-\frac{x^{2}+y^{2}}{R(z)^{2}}+\frac{2\lambda}{R(z)}\right)
\nonumber \\
&=& 2v_{l}\frac{R(l)^{2}}{R(z)^{2}}\left(1-\frac{r^{2}}{R(z)^{2}}+\frac{2\lambda}{R(z)}\right),
\label{eq:Sv13}
\end{eqnarray}
where $l$ represents the meniscus position and $v_{l}$ is a characteristic velocity of the flow front at $z=l$.  In fact, such a parabolic velocity profile in a conical tube can be confirmed by the microscopic molecular dynamic simulation~\cite{Mondal2021}. In axially symmetric tubes with strongly corrugated wall, however, the existence of a closed eddy in a pocket of wall is suggested theoretically~\cite{Scholle2004,Marner2017}.  Even in simple conical tubes, a similar closed eddy is observed by numerical simulation when the Reynolds number is large~\cite{Goli2022}.  In such cases when the wall corrugation is strong or the Reynolds number is large, our hydrodynamic model based on the parabolic velocity profile will be unrealistic.

From Eq~(\ref{eq:Sv13}), the volumetric flow rate $Q(z)$ across the cross section
\begin{equation}
\int\int dx dy=\int_{0}^{2\pi}d\theta\int_{0}^{R(z)}rdr=\pi R(z)^{2},
\label{eq:Sv14}
\end{equation}
at $z$ becomes
\begin{equation}
Q(z)=\int\int v_{z}(z,r,t)dx dy=\pi R(l)^{2}\left(1+\frac{4\lambda}{R(z)}\right)v_{l}.
\label{eq:Sv15}
\end{equation}
Suppose we consider a steady flow, this volumetric flow rate must be constant [$Q(z)=Q$] due to the mass conservation.  Therefore, the slip length $\lambda$ must be proportional to the radius $R(z)$ so that
\begin{equation}
\lambda=\lambda_{1}R(z),
\label{eq:Sv16}
\end{equation}
and the proportionality constant $\lambda_{1}$ plays a role of a slip length in a tube with axial variation.  Since we assume a quasi-steady flow, we continue to use Eq.~(\ref{eq:Sv16}).  It is a logical consequence of the mathematical form of Navier's slip length.  The magnitude of $\lambda_{1}$ will be determined from the material properties such as Young's contact angle of wettability~\cite{Wu2017c}.  Equation~(\ref{eq:Sv16}) has been already derived for diverging conical tubes by Tran-Duc {\it et al.}~\cite{Tran-Duc2019} from the exact solution of the Stokes equation using a spherical coordinate system.  Our result in Eq.~(\ref{eq:Sv16}) is more general and applicable to any tubes with varying radius $R(z)$ including conical tubes. Then, the volumetric flow rate becomes
\begin{equation}
Q=\pi R(l)^{2}\left(1+4\lambda_{1}\right)v_{l},
\label{eq:Sv17}
\end{equation}
which does not depend on the position $z$. The average velocity $dl/dt$ of the meniscus at $z=l$ can be defined by
\begin{equation}
Q=\pi R(l)^{2}\frac{dl}{dt}.
\label{eq:Sv18}
\end{equation}
and is expressed by the characteristic velocity $v_{l}$ by
\begin{equation}
\frac{dl}{dt}=\left(1+4\lambda_{1}\right)v_{l}.
\label{eq:Sv19}
\end{equation}
from Eq.~(\ref{eq:Sv17}).

From Eq.~(\ref{eq:Sv16}), the velocity profiles $v_{z}$ in Eq.~(\ref{eq:Sv13}) become
\begin{equation}
v_{z}\left(z,r,t\right) = 2v_{l}\frac{R(l)^{2}}{R(z)^{2}}\left(1+2\lambda_{1}-\frac{r^{2}}{R(z)^{2}}\right),
\label{eq:Sv20}
\end{equation}
and the radial components $v_{x}\left(z,r,t\right)$ and  $v_{y}\left(z,r,t\right)$ can be constructed from Eq.~(\ref{eq:Sv20})~\cite{Liou2009,Wang2013,Nissan2016} to satisfy the continuity equation in Eq.~(\ref{eq:Sv5}).  Since we look at the flow rate $Q$, we consider only the axial component $v_{z}\left(z,r,t\right)$.  Then, the left-hand side of Eq.~(\ref{eq:Sv9}) can be calculated and written as
\begin{equation}
\mu\left( \frac{\partial^{2}v_{z}}{\partial x^{2}}  + \frac{\partial^{2}v_{z}}{\partial y^{2}} + \frac{\partial^{2}v_{z}}{\partial z^{2}} \right)
=-8\mu v_{l}R(l)^{2}\frac{s(z,r)}{R(z)^4},
\label{eq:Sv21}
\end{equation}
with
\begin{eqnarray}
s(z,r)&=&\left[1-\frac{1}{2}\left(3\left(1+2\lambda_{1}\right)-\frac{10 r^{2}}{R(z)^2}\right)\left(\frac{dR}{dz}\right)^{2}+\right.
\nonumber \\
&+&\left. \frac{1}{2}\left(\left(1+2\lambda_{1}\right)-\frac{2r^{2}}{R(z)^{2}}\right)R(z)\frac{d^{2}R}{dz^{2}}\right].
\label{eq:Sv22}
\end{eqnarray}
The second term proportional to $(dR/dz)^{2}$ and the third term proportional to $d^{2}R/dz^{2}$ are the perturbation corrections due to the effect of the non-slowly-varying radius $R(z)$, and $s(z,r)=1$ for straight cylindrical tubes.

The Stokes equation in Eq.~(\ref{eq:Sv9}) is written as
\begin{equation}
-8\mu v_{l}R(l)^{2}\frac{s(z,r)}{R(z)^4}=\frac{\partial p(z,r)}{\partial z},
\label{eq:Sv23}
\end{equation}
which can be combined with Eq.~(\ref{eq:Sv17}) and we have
\begin{equation}
Q=\frac{\pi R(z)^{4}\left(1+4\lambda_{1}\right)}{8\mu s(z,r)}\left(-\frac{\partial p(z,r)}{\partial z}\right),
\label{eq:Sv24}
\end{equation}
which is a generalization of the original Hagen-Poiseulle law for a straight cylindrical tube in Eq.~(\ref{eq:Sv1}) to an axially symmetric tube.

Equation (\ref{eq:Sv24}) clearly suggests that the pressure $p(z,r)$ depends on the radial coordinate $r$, which was also found for a divergent conical tube from the exact solution of the Stokes equation~\cite{Tran-Duc2019}. Here we introduce the averaging over the radial coordinate defined by
\begin{equation}
\bar{f}(z)=\frac{\int\int dxdy f(z,r)}{\int\int dxdy}=\frac{\int_{0}^{R(z)}2\pi r dr f(z,r)}{\pi R(z)^{2}},
\label{eq:Sv25}
\end{equation}
where $f$ is a general function of $z$ and $r$, since we are interested in the axial $z$ component to know the flow rate $Q$.  Then, Eq.~(\ref{eq:Sv23}) can be averaged and it becomes
\begin{equation}
-8\mu v_{l}R(l)^{2}\frac{\bar{s}(z)}{R(z)^4}=\frac{d\bar{p}(z)}{dz}
\label{eq:Sv26}
\end{equation}
with
\begin{equation}
\bar{s}(z) = \left[1+\left(1-3\lambda_{1}\right)\left(\frac{dR}{dz}\right)^{2} + \lambda_{1}R(z)\frac{d^{2}R}{dz^{2}}\right],
\label{eq:Sv27}
\end{equation}
which can be combined with Eq.~(\ref{eq:Sv17}), and the volumetric flow rate in Eq.~(\ref{eq:Sv24}) is further simplified to
\begin{equation}
Q=\frac{\pi R(z)^{4}\left(1+4\lambda_{1}\right)}{8\mu \bar{s}(z)}\left(-\frac{d \bar{p}(z)}{d z}\right).
\label{eq:Sv28}
\end{equation}
The Hagen-Poiseulle law in Eq.~(\ref{eq:Sv1}) is recovered for straight cylindrical tubes [$R(z)=$constant, $\bar{s}(z)=1$] with no slip [$\lambda_{1}=0$].  The extension of Eq.~(\ref{eq:Sv1}) with a slip is simply given by setting $\bar{s}(z)=1$ and $\lambda_{1}\neq 0$ in Eq.~(\ref{eq:Sv28})~\cite{Washburn1921,Wu2017c,Suk2017,Kavokine2021,Willmott2011}.

Equation~(\ref{eq:Sv26}) can be integrated, and we obtain
\begin{equation}
-8\mu v_{l}R(l)^{2}\int_{0}^{l}\frac{\bar{s}(z)}{R(z)^4}dz=\bar{p}(l)-\bar{p}(0),
\label{eq:Sv29}
\end{equation}
or
\begin{equation}
\bar{p}(0)-\bar{p}(l)=\frac{8\mu Q}{\pi\left(1+4\lambda_{1}\right)}\int_{0}^{l}\frac{\bar{s}(z)}{R(z)^4}dz
\label{eq:Sv30},
\end{equation}
from Eq.~(\ref{eq:Sv17}). Equation (\ref{eq:Sv30}) is a generalization of Eq.~(\ref{eq:Sv3}) for a tube with a non-slowly-varying radius $R(z)$.

Now, we consider a quasi-steady capillary flow of spontaneous imbibition driven by a surface tension $\gamma$, and
 approximate the pressure difference by the Laplace formula
\begin{equation}
\bar{p}(l)-\bar{p}(0)\simeq -\frac{C}{R(l)},
\label{eq:Sv31}
\end{equation}
where $C=2\gamma\cos\left(\theta_{\rm Y}+\phi(l)\right)$, $\theta_{\rm Y}$ is Young's contact angle, and the tilt angle $\phi(l)$ of the tube wall at $l$ is defined by $\tan \phi (l)=dR/dl$.  The spontaneous imbibition occurs when $C$ is positive.  Hence, Young's contact angle must satisfy $\theta_{\rm Y}+\phi (l)<90^{\circ}$.

We follow the previous researchers and assume that $C$ is approximately constant~\cite{Sharma1991,Staples2002,Young2004,Reyssat2008,Urteaga2013,Berli2014,Elizalde2014,Gorce2016,Khatoon2020,Tran-Duc2020,Iwamatsu2022}.  In fact, the validity of this assumption can be checked mathematically from  $\cos\left(\theta_{\rm Y}+\phi\right)\sim 1-(1/2)(dR/dz)^{2}$ for a highly hydrophilic wall $\theta_{\rm Y}\sim 0^{\circ}$~\cite{Reyssat2008}.  Though this correction is similar to that in Eq.~(\ref{eq:Sv27}), we adopt this assumption ($C=$constant) because it is convenient to seek for an analytical formulation; otherwise, we resort to only a fully numerical case-by-case calculation.  Then, Eq.~(\ref{eq:Sv29}) can be combined with Eq.~(\ref{eq:Sv19}), and the differential equation for the meniscus position $l$ can be written as
\begin{equation}
\frac{dl}{dt}=\frac{C\left(1+4\lambda_{1}\right)}{8\mu\left[C_{1}(l)+\left(1-3\lambda_{1}\right)C_{2}(l)+\lambda_{1}C_{3}(l)\right]},
\label{eq:Sv32}
\end{equation}
where
\begin{eqnarray}
C_{1}(l) &=& R(l)^{3}\int_{0}^{l}\frac{1}{R(z)^4}dz,
\label{eq:Sv33} \\
C_{2}(l) &=& R(l)^{3}\int_{0}^{l}\frac{1}{R(z)^4}\left(\frac{dR}{dz}\right)^{2}dz,
\label{eq:Sv34} \\
C_{3}(l) &=& R(l)^{3}\int_{0}^{l}\frac{1}{R(z)^3}\left(\frac{d^{2}R}{dz^{2}}\right)dz,
\label{eq:Sv35}
\end{eqnarray}
and the non-slowly-varying perturbation corrections are expressed by the integrals $C_{2}$ proportional to $(dR/dz)^{2}$ and $C_{3}$ proportional to $d^{2}R/dz^{2}$. Also, the effect of the boundary slip $\lambda_{1}$ is always coupled to the perturbations $C_{2}$ and $C_{3}$.  In straight cylindrical tubes [$R(z)=$constant, $\bar{s}(z)=1$], both $C_{2}$ and $C_{3}$ vanish, and the volumetric flow rate $Q$ is given by Eq.~(\ref{eq:Sv28}) with $\bar{s}(z)=1$, which is the Hagen-Poiseulle formula with the non-zero slip boundary condition.~\cite{Washburn1921,Wu2017c,Suk2017,Kavokine2021,Willmott2011}.   In Sec. III, we will study typical geometries considered previously to see when these two contributions can be negligible.

\section{\label{sec:sec3}Application to typical geometries }

\subsection{Capillary imbibition in a diverging and a converging conical tube}

\begin{figure}[htbp]
\begin{center}
\includegraphics[width=0.6\linewidth]{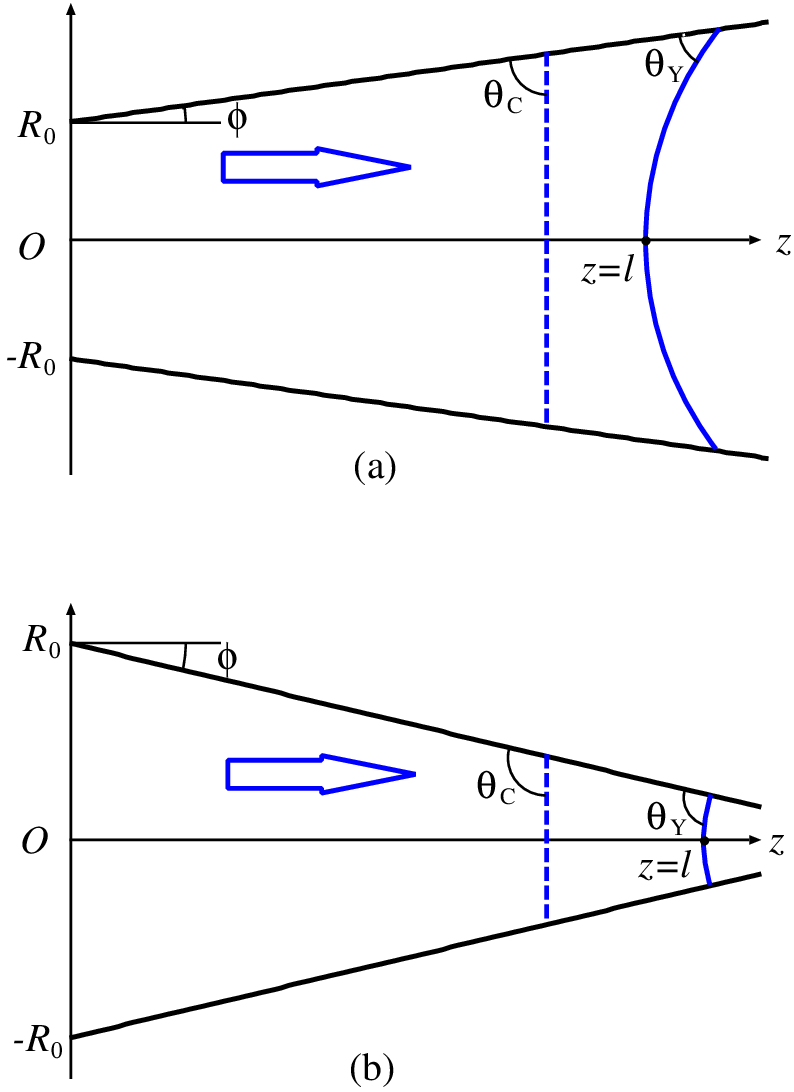}
\end{center}
\caption{
(a) A diverging and (b) a converging conical tube.  The flow direction is from left to right as indicated by an arrow.  The position of meniscus is at $z=l$ and the contact angle $\theta_{\rm Y}$ satisfies the condition $\theta_{\rm Y}<\theta_{\rm c}=90^{\circ}-\phi$ for a diverging tube and  $\theta_{\rm Y}<\theta_{\rm c}=90^{\circ}+\phi$ for a converging tube with $\phi>0$ being the tilt angle of the conical wall.
}
\label{fig:SL1}
\end{figure}

As the first example of an axially symmetric tube, we consider a conical tube (Fig.~\ref{fig:SL1}) whose radius changes according to
\begin{equation}
R(z) = R_{0} + \alpha z,
\label{eq:Sv36}
\end{equation}
where $\alpha<0$ corresponds to a converging and $\alpha>0$ to a diverging tube.

This simple conical tube acts as a liquid diode, namely, the converging direction acts as the forward and the diverging direction acts as the reverse direction of the electronic diode device~\cite{Comanns2015,Singh2020,Panter2020,Iwamatsu2022,Xu2023, Iwamatsu2023}.  Because spontaneous imbibition is possible only when the contact angle $\theta_{\rm Y}$ satisfies $\theta_{\rm Y}<\theta_{\rm c}$, where $\theta_{\rm c}$ is a critical angle shown in Fig.~\ref{fig:SL1}, the spontaneous one-way transport toward the converging direction occurs when the contact angle $\theta_{\rm Y}$ satisfies $90^{\circ}+\phi>\theta_{\rm Y}>90^{\circ}-\phi$ since $\theta_{\rm c}=90^{\circ}+\phi$ in a converging [Fig.~\ref{fig:SL1}(a)] and  $\theta_{\rm c}=90^{\circ}-\phi$ in a diverging [Fig.~\ref{fig:SL1}(b)] tube with $\phi$ being the tilt angle of the wall ($\alpha=\pm \tan\phi$).  Therefore, a large tilt angle $\phi$ and, therefore, a non-slowly-varying conical tube is more advantageous to realize a liquid diode~\cite{Iwamatsu2022,Iwamatsu2023}.

Here, we consider the transport to both directions so that we consider a wetting liquid with 
\begin{equation}
\theta_{\rm Y}<90^{\circ}-\phi=90^{\circ}-\arctan \left|\frac{dR}{dz}\right|, 
\label{eq:Sv37}
\end{equation}
with $dR/dz=\alpha$ so that the spontaneous imbibition occurs by the action of capillary pressure in Eq.~(\ref{eq:Sv31}).  Since $d^{2}R/dz^{2}=0$, Eqs.~(\ref{eq:Sv34}) and (\ref{eq:Sv35}) become
\begin{equation}
C_{2}(l)=\alpha^{2}C_{1}(l),\;\;\;C_{3}(l)=0,
\label{eq:Sv38}
\end{equation}
and Eq.~(\ref{eq:Sv32}) is written as
\begin{equation}
\frac{dl}{dt}=\frac{C\left(1+4\lambda_{1}\right)}{8\mu\left[1+\left(1-3\lambda_{1}\right)\alpha^{2}\right]C_{1}(l)},
\label{eq:Sv39}
\end{equation}
where
\begin{eqnarray}
C_{1}(l) &=& R(l)^{3}\int_{0}^{l}\frac{1}{R(z)^4}dz,
\nonumber \\
 &=& \left(R_{0} + \alpha l\right)^{3}\int_{0}^{l}\left(R_{0} + \alpha z\right)^{-4}dz,
\nonumber \\
&=& \frac{1}{3\alpha}\left[\left(1+\frac{\alpha}{R_{0}}l\right)^{3}-1\right].
\label{eq:Sv40}
\end{eqnarray}
Therefore, the non-slowly-varying perturbation parameter is $\alpha^{2}$ and is absorbed into the denominator of Eq.~(\ref{eq:Sv39}), which is written as
\begin{equation}
\frac{dl}{dt}\left[\left(1+\frac{\alpha}{R_{0}}l\right)^{3}-1\right]=\frac{3C\left(1+4\lambda_{1}\right)\alpha}
{8\mu\left[1+\left(1-3\lambda_{1}\right)\alpha^{2}\right]}.
\label{eq:Sv41}
\end{equation}
By introducing non-dimensional variables $L$ and $T$ instead of $l$ and $t$ defined by
\begin{equation}
L=\frac{\left|\alpha\right|}{R_{0}}l,\;\;\;T=\frac{C\left(1+4\lambda_{1}\right)\alpha^{2}}{\mu R_{0}\left[1+\left(1-3\lambda_{1}\right)\alpha^{2}\right]}t,
\label{eq:Sv42}
\end{equation}
Equation (\ref{eq:Sv41}) is transformed into
\begin{equation}
\frac{dL}{dT}\left(\left(1\pm L\right)^{3}-1\right)=\pm\frac{3}{8},
\label{eq:Sv43}
\end{equation}
where the sign $\pm$ denotes the sign of $\alpha$.  Equation (\ref{eq:Sv43}) has been derived from Eq.~(\ref{eq:Sv1}) and used to derive various scaling laws of imbibition~\cite{Reyssat2008,Urteaga2013,Iwamatsu2022}.  Therefore, the perturbation of a non-slowly-varying radius $R(z)$ is included only in the scaling for $T$ in Eq.~(\ref{eq:Sv42}) by the term proportional to $\alpha^{2}=(dR/dZ)^{2}$ in the denominator.

Equation (\ref{eq:Sv43}) can be integrated and we have
\begin{equation}
T = 4L^{2}\pm\frac{8}{3}L^{3} + \frac{2}{3}L^{4}
\label{eq:Sv44}
\end{equation}
which gives the time $T_{\rm D}$ necessary to reach the position $L_{\rm D}$ of a diverging tube with an inlet radius $R_{0}=R_{\rm D}$ and $T_{\rm C}$ to reach $L_{\rm C}$ of a converging tube with an inlet radius $R_{0}=R_{\rm C}$
\begin{eqnarray}
T_{\rm D} &=& 4L_{\rm D}^{2}+\frac{8}{3}L_{\rm D}^{3}+\frac{2}{3}L_{\rm D}^{4},\;\;\;(\alpha>0,\mbox{Diverging}),
\label{eq:Sv45} \\
T_{\rm C} &=& 4L_{\rm C}^{2}-\frac{8}{3}L_{\rm C}^{3}+\frac{2}{3}L_{\rm C}^{4},\;\;\;(\alpha<0,\mbox{Converging}).
\label{eq:Sv46}
\end{eqnarray} 
They were originally obtained by Urteaga et al.~\cite{Urteaga2013} [Eqs.~(3) and (4)].  Note that $L_{\rm C}\le 1$ since $R_{\rm C}-|\alpha| l \ge 0$ for a converging tube.

\begin{figure}[htbp]
\begin{center}
\includegraphics[width=0.9\linewidth]{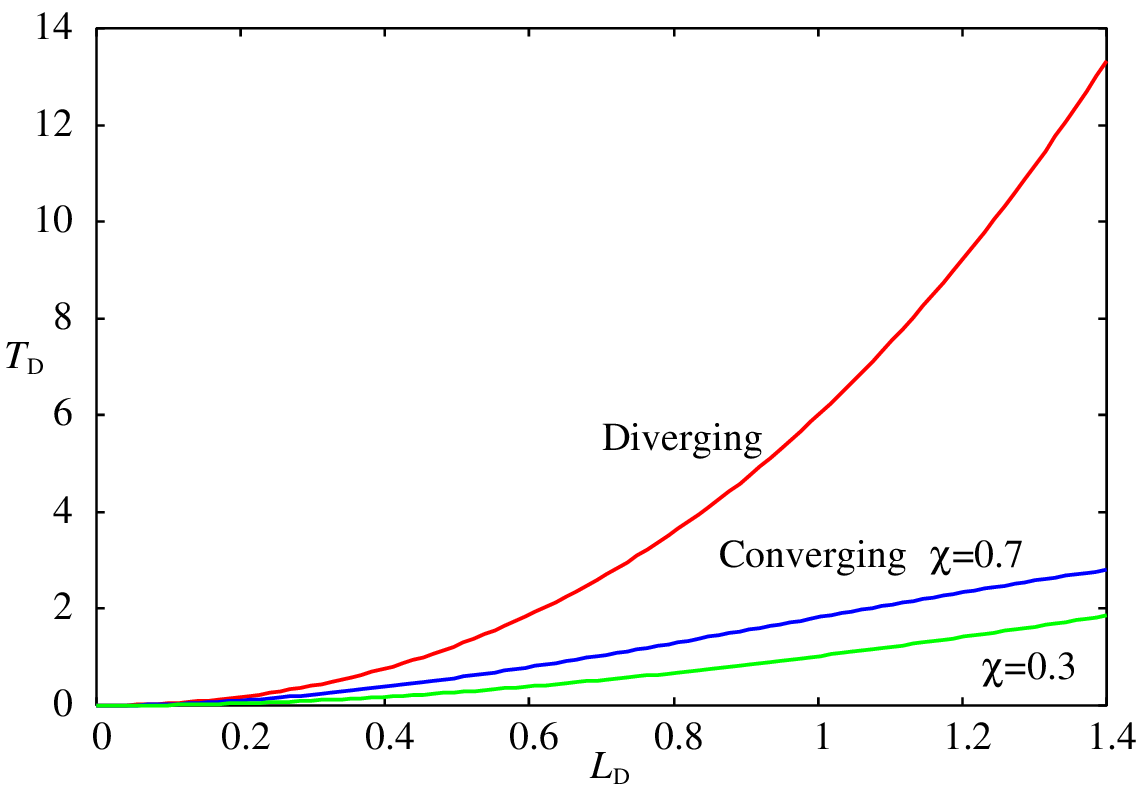}
\end{center}
\caption{
The time $T_{\rm D}$ necessary to reach $L_{\rm D}$ of a diverging conical tube and that of a converging tube of the same shape when $\chi=R_{\rm D}/R_{\rm C}$=0.3 and 0.7. 
}
\label{fig:SL2}
\end{figure}

To compare the performance of a diverging and a converging conical tube of the {\it same shape} with the inlet radius $R_{\rm D}$ and the outlet radius $R_{\rm C}$ ($R_{\rm D}<R_{\rm C}$), we note
\begin{equation}
L_{\rm C}=\frac{R_{\rm D}}{R_{\rm C}}L_{\rm D}=\chi L_{\rm D},\;\;\;T_{\rm C}=\frac{R_{\rm D}}{R_{\rm C}}T_{\rm D}=\chi T_{\rm D},\;\;\;\chi<1,
\label{eq:Sv47}
\end{equation}
where $\chi=R_{\rm D}/R_{\rm C}<1$ is the asymmetry of the conical tube.  Then, the performance of the diverging conical tube in Eq.~(\ref{eq:Sv45}) should be compared not with the performance of the converging tube in Eq.~(\ref{eq:Sv46}) but with
\begin{equation}
T_{\rm D} = 4\chi L_{\rm D}^{2}-\frac{8}{3}\chi^{2}L_{\rm D}^{3}+\frac{2}{3}\chi^{3}L_{\rm D}^{4},\;\;\;(\alpha<0,\mbox{Converging})
\label{eq:Sv48}
\end{equation}
derived from Eq.~(\ref{eq:Sv46}) by the transformation in Eq.~(\ref{eq:Sv47}).  Since $L_{\rm C}\le 1$, $L_{\rm D}=L_{\rm C}/\chi\le 1/\chi$ for a converging tube..

In Fig.~\ref{fig:SL2}, we compare the time $T_{\rm D}$ that is necessary to reach the position $L_{\rm D}$ of a diverging conical tube and that of a converging tube, where we use the scaling in Eq.~(\ref{eq:Sv42}) with the inlet radius $R_{0}=R_{\rm D}$ of a diverging cone for both directions.  Apparently, the flow in a converging tube is faster than that in a diverging tube, which was already pointed out~\cite{Urteaga2013,Iwamatsu2022}.  Here, we show that this conclusion is generic and does not depend on whether the radius of the tube is slowly-varying or not.

\subsection{Capillary imbibition in a diverging power-law-shaped tube}

\begin{figure}[htbp]
\begin{center}
\includegraphics[width=0.6\linewidth]{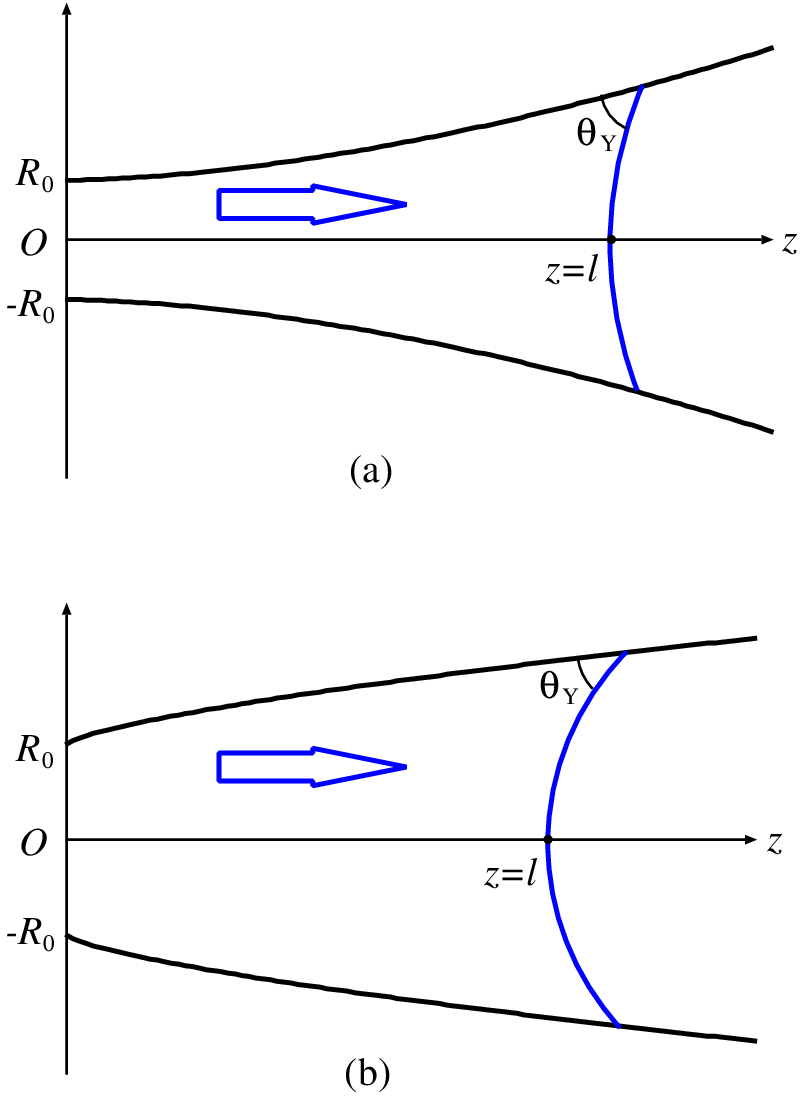}
\end{center}
\caption{
(a) A diverging power-law-shaped tube with the exponent $n>1$ and (b) that with the exponent $1>n>0$.
}
\label{fig:SL3}
\end{figure}

Next, we consider the general case of power-law-shaped diverging tubes shown in Fig.~\ref{fig:SL3}, whose radius changes according to  
\begin{equation}
R(z) = R_{0} + \alpha z^{n},\;\;\;(\alpha>0),
\label{eq:Sv49}
\end{equation}
where the dimension of $\alpha$ depends on the exponent $n>0$. Reyssat et al.~\cite{Reyssat2008} considered this problem and derived classical diffusive scaling law $l \propto t^{1/2}$ at an early stage and $l\propto t^{1/(3n+1)}$ at a late stage.  Apparently, the slowly-varying condition in Eq.~(\ref{eq:Sv1}) will be violated as $dR/dz\propto z^{n-1}\rightarrow \infty$ at a late stage when $n>1$ [Fig.~\ref{fig:SL3}(a)].  Therefore, it is interesting to see if the effect of non-slowly-varying radius $R(z)$ affects these scaling rules.

By introducing non-dimensional variables similar to those in Eq.~(\ref{eq:Sv42}) defined by
\begin{eqnarray}
L=\left(\frac{\alpha}{R_{0}}\right)^{1/n}h,\;\;\;Z=\left(\frac{\alpha}{R_{0}}\right)^{1/n}z,
\nonumber \\
T=\frac{C\left(1+4\lambda_{1}\right) R_{0}}{\mu}\left(\frac{\alpha}{R_{0}}\right)^{2/n}t,
\label{eq:Sv50}
\end{eqnarray}
Equations.~(\ref{eq:Sv33}) to (\ref{eq:Sv35}) will be written as
\begin{eqnarray}
C_{1}(L) &=& R_{0}^{-1}\left(\frac{R_{0}}{\alpha}\right)^{1/n}
\left(1+ L^{n}\right)^{3}\sigma_{1}(L),
\label{eq:Sv51} \\
C_{2}(L) &=& \epsilon_{\rm D} R_{0}^{-1}\left(\frac{R_{0}}{\alpha}\right)^{1/n}
n^{2} \left(1+ L^{n}\right)^{3}\sigma_{2}(L),
\label{eq:Sv52} \\
C_{3}(L) &=& \epsilon_{\rm D} R_{0}^{-1}\left(\frac{R_{0}}{\alpha}\right)^{1/n}
n(n-1) \left(1+ L^{n}\right)^{3}\sigma_{3}(L),
\label{eq:Sv53}
\end{eqnarray}
where
\begin{eqnarray}
\sigma_{1}(L) &=& \int_{0}^{L}\left(1+ Z^{n}\right)^{-4}dZ,
\label{eq:Sv54} \\
\sigma_{2}(L) &=& \int_{0}^{L}\left(1+ Z^{n}\right)^{-4}Z^{2n-2}dZ,
\label{eq:Sv55} \\
\sigma_{3}(L) &=&\int_{0}^{L}\left(1+ Z^{n}\right)^{-3}Z^{n-2}dZ,
\label{eq:Sv56}
\end{eqnarray}
and the non-dimensional perturbation parameter $\epsilon_{\rm D}$ for the diverging tube is given by
\begin{equation}
\epsilon_{\rm D}=\left(\frac{\alpha R_{0}^{n}}{R_{0}}\right)^{2/n}=\left(\frac{\mbox{increment of radius}}{\mbox{inlet radius}}\right)^{2/n}.
\label{eq:Sv57}
\end{equation}
Therefore, the non-slowly-varying effect is determined from how rapidly the radius changes near the inlet.  Then, Eq.~(\ref{eq:Sv32}) becomes
\begin{eqnarray}
\frac{dL}{dT}\left(1+ L^{n}\right)^{3}\left[\sigma_{1}(L) + \epsilon_{\rm D}\left(1-3\lambda_{1}\right)n^{2} \sigma_{2}(L)\right.
\nonumber \\
\left. + \epsilon_{\rm D}\lambda_{1}n(n-1) \sigma_{3}(L)\right]=\frac{1}{8}
\label{eq:Sv58}
\end{eqnarray}
which determines the dynamics of the capillary flow in a diverging power-law-shaped tube when its radius is not necessarily slowly-varying.

It is straightforward to consider the asymptotic limit~\cite{Reyssat2008} of Eq.~(\ref{eq:Sv58}).  In an early stage $L\ll 1$, we have
\begin{eqnarray}
\sigma_{1}(L) &\simeq& L,
\label{eq:Sv59} \\
\sigma_{2}(L) &\simeq& L^{2n-1},
\label{eq:Sv60} \\
\sigma_{3}(L) &\simeq& L^{n-1}.
\label{eq:Sv61}
\end{eqnarray}
Therefore, Eq.~(\ref{eq:Sv58}) becomes
\begin{equation}
 \frac{dL}{dT}\left[L + \epsilon_{\rm D}\left(1-3\lambda_{1}\right)n^{2} L^{2n-1}
 +\epsilon_{\rm D}\lambda_{1}n(n-1) L^{n-1}\right]=\frac{1}{8},
 \label{eq:Sv62}
\end{equation}
in an early stage.

If the non-slip boundary condition is applicable ($\lambda_{1}=0$), the last term of the left-hand side of Eq.~(\ref{eq:Sv62}) disappears.  When $n\ge 1$, the first term Eq.~(\ref{eq:Sv62}) is dominant and we recover the classical diffusive law
\begin{equation}
\frac{dL}{dT}L\sim\frac{1}{8}\rightarrow L\propto T^{1/2},\;\;\;(\lambda_{1}=0,n\ge 1)
\label{eq:Sv63}
\end{equation}
derived by Reyssat {\it et al.}~\cite{Reyssat2008}  When $1>n>0$, no matter how small the perturbation parameter $\epsilon_{\rm D}$ is, the second term of the left-hand side of Eq.~(\ref{eq:Sv62}) becomes dominant in an early stage and we have a new scaling rule
\begin{equation}
\frac{dL}{dT}L^{2n-1}\sim\frac{1}{8}\rightarrow L\propto T^{1/2n},\;\;\;(\lambda_{1}=0,1>n>0)
\label{eq:Sv64}
\end{equation}
from the non-slowly-varying radius $R(z)$.

If the slip length cannot be negligible ($\lambda \neq 0$) and $n\ge 2$, no matter how large the perturbation parameter $\epsilon_{\rm D}$ is, we can neglect the last two terms of the left-hand side of Eq.~(\ref{eq:Sv62}) and we find a classical diffusion law again:
\begin{equation}
\frac{dL}{dT}L\sim\frac{1}{8}\rightarrow L\propto T^{1/2},\;\;\;(\lambda_{1}\neq 0, n\ge 2).
\label{eq:Sv65}
\end{equation}
When $2\ge n > 0$, on the other hand, the dominant term becomes the last term in Eq.~(\ref{eq:Sv62}) and we have a new scaling rule:
\begin{equation}
\frac{dL}{dT}L^{n-1}\sim \frac{1}{8}\rightarrow L\propto T^{1/n},\;\;\;(\lambda_{1}\neq 0, 2\ge n>0).
\label{eq:Sv66}
\end{equation}
Therefore, a non-slowly-varying radius affects an early-stage scaling rule when $\lambda_{1}=0$ and $1>n>0$, and when $\lambda_{1}\neq 0$ and $2\ge n >0$, and the scaling rules are given by Eqs.~(\ref{eq:Sv64}) and (\ref{eq:Sv66}), respectively.  They are different from the classical diffusive law in Eqs.~(\ref{eq:Sv63}) and (\ref{eq:Sv65}) derived by Reyssat {\it et al.}~\cite{Reyssat2008}

Similarly, at a late stage $L\gg 1$, Eq.~(\ref{eq:Sv58}) can be approximately written as
\begin{eqnarray}
\frac{dL}{dT}\left(1+ L^{n}\right)^{3}\left[\sigma_{1}(\infty) + \epsilon_{\rm D}\left(1-3\lambda_{1}\right)n^{2} \sigma_{2}(\infty)\right.
\nonumber \\
+ \epsilon_{\rm D}\left. \lambda_{1}n(n-1) \sigma_{3}(\infty)\right]=\frac{1}{8}
\label{eq:Sv67}
\end{eqnarray}
because $\sigma_{1}(\infty)$, $\sigma_{2}(\infty)$, and $\sigma_{3}(\infty)$ are all finite.  Therefore, at a late stage
\begin{equation}
\frac{dL}{dT}L^{3n}\sim \frac{1}{8}\rightarrow L\propto T^{\frac{1}{3n+1}},
\label{eq:Sv68}
\end{equation}
which has been found previously by Reyssat {\it et al.}~\cite{Reyssat2008} using the slowly-varying approximation in Eq.~(\ref{eq:Sv1}).  As far as the asymptotic limit at a late stage is concerned, the effect of perturbation $\epsilon_{\rm D}$ does not affect the scaling rule in Eq.~(\ref{eq:Sv68}).   Therefore, even if the radius is not necessarily slowly varying, the power-law-shaped tube shows the same scaling rule at a late stage.

\subsection{Capillary imbibition in a converging power-law-shaped tube}

\begin{figure}[htbp]
\begin{center}
\includegraphics[width=0.6\linewidth]{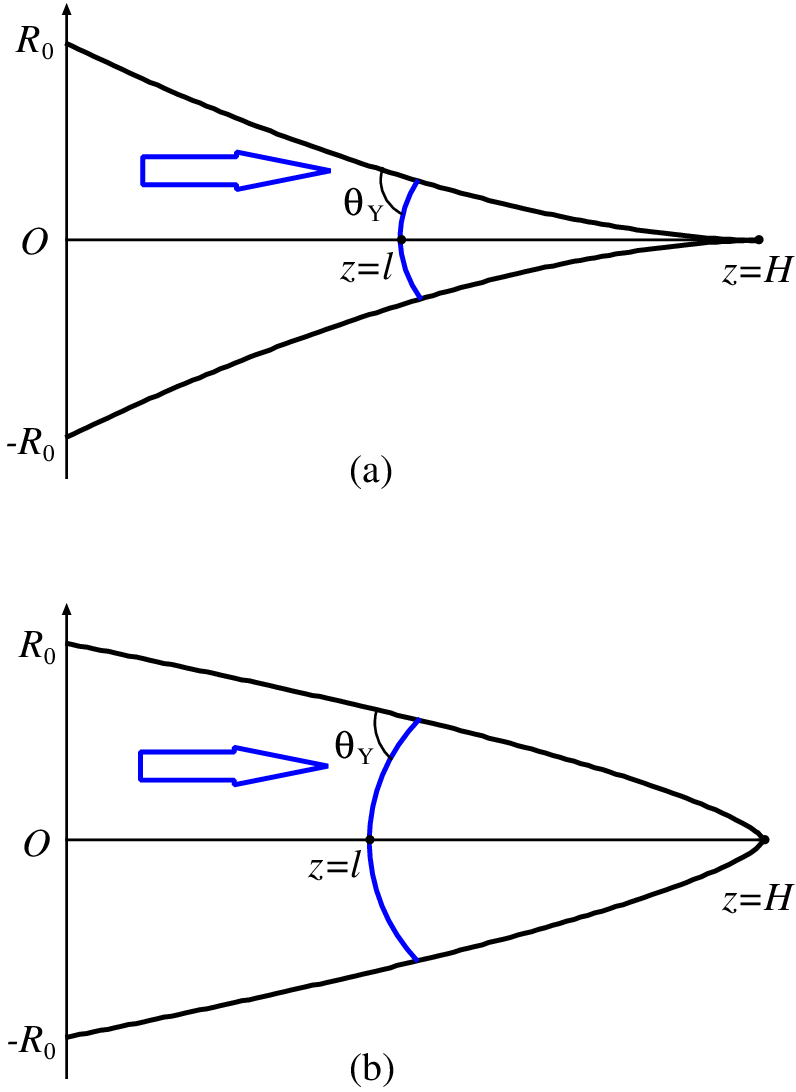}
\end{center}
\caption{
(a) A converging power-law-shaped tube with the exponent $n>1$ and (b) that with the exponent $1>n>0$.
}
\label{fig:SL4}
\end{figure}

Gorce {\it et al.}~\cite{Gorce2016} have studied converging power-law-shaped tubes whose radius changes as
\begin{equation}
R(z)=R_{0}\left(\frac{H-z}{H}\right)^{n},\;\;\;0\le z\le H,
\label{eq:Sv69}
\end{equation}
where $H$ is the length of the converging tube.  Typical shapes are shown in Fig.~\ref{fig:SL4} where we show the tube shape for $n>1$ and $1>n>0$.  It is assumed that there is a small hole at the outlet at $z=H$ so that the fluid can flow out from the tube and any divergence associated with $R(z)\rightarrow 0$ as $z\rightarrow H$ is neglected.  They~\cite{Gorce2016} showed that the flow rate can be faster than that in a straight cylindrical tube beating classical Washburn's law~\cite{Washburn1921}.  Furthermore, there exists the optimal index $n_{\rm min}$ for which the imbibition time is shortest.  We will see how their results~\cite{Gorce2016} will be modified by the non-slowly-varying effect.

By introducing the non-dimensional variables
\begin{equation}
L=\frac{l}{H},\;\;\;Z=\frac{z}{H},\;\;\;T=\frac{C\left(1+4\lambda_{1}\right) R_{0}}{\mu H^{2}}t,
\label{eq:Sv70}
\end{equation}
Equation~(\ref{eq:Sv33}) becomes
\begin{eqnarray}
C_{1}(L) &=& HR_{0}^{-1}\left(1-L\right)^{3n}\int_{0}^{L}\left(1-Z\right)^{-4n}dZ 
\nonumber \\
&=& \frac{HR_{0}^{-1}}{4n-1}\left(\left(1-L\right)^{1-n}-\left(1-L\right)^{3n}\right).
\label{eq:Sv71}
\end{eqnarray}
Similarly, Eqs.~(\ref{eq:Sv34}) and (\ref{eq:Sv35}) become
\begin{eqnarray}
C_{2}(L) &=& \epsilon_{\rm C}\frac{n^{2}HR_{0}^{-1}}{2n+1}\left(\left(1-L\right)^{n-1}-\left(1-L\right)^{3n}\right),
\label{eq:Sv72} \\
C_{3}(L) &=& \epsilon_{\rm C}\frac{n(n-1)HR_{0}^{-1}}{2n+1}\left(\left(1-L\right)^{n-1}-\left(1-L\right)^{3n}\right),
\label{eq:Sv73}
\end{eqnarray}
where the perturbation parameter $\epsilon_{\rm C}$ for the converging tube is defined by
\begin{equation}
\epsilon_{\rm C}=\left(\frac{R_{0}}{H}\right)^{2},
\label{eq:Sv74}
\end{equation}
whose meaning is apparent.

Using these non-dimensional variables in Eq.~(\ref{eq:Sv70}), Eq.~(\ref{eq:Sv32}) is written as
\begin{eqnarray}
&&\frac{dL}{dT}\left[\frac{1}{4n-1}\left\{\left(1-L\right)^{1-n}-\left(1-L\right)^{3n}\right\}\right.
\nonumber \\
&&\left. +\epsilon_{\rm C}\left(\frac{n^{2}}{2n+1}-\lambda_{1}n\right)\left\{\left(1-L\right)^{1-n}-\left(1-L\right)^{3n}\right\}\right]
\nonumber \\
&&=\frac{1}{8}
\label{eq:Sv75}
\end{eqnarray}
which can be integrated and the time $T$ necessary to reach $L$ is given by
\begin{eqnarray}
T&=&T_{\rm end}+\frac{8}{4n-1}\left\{ \frac{\left(1-L\right)^{3n+1}}{3n+1}-\frac{\left(1-L\right)^{2-n}}{2-n}\right\}
\nonumber \\
&+& 8\epsilon\left\{\frac{\left(1-L\right)^{3n+1}}{3n+1} - \frac{\left(1-L\right)^{n}}{n} \right\}
\label{eq:Sv76}
\end{eqnarray}
with
\begin{equation}
\epsilon=\left(\frac{n^{2}}{2n+1} - \lambda_{1}n\right)\epsilon_{\rm C},
\label{eq:Sv77}
\end{equation}
is a new perturbation parameter, and
\begin{equation}
T_{\rm end}=\frac{8}{\left(2-n\right)\left(3n+1\right)}+8\epsilon\frac{2n+1}{n\left(3n+1\right)},
\label{eq:Sv78}
\end{equation}
is the time necessary to reach the outlet at $z=H (L=1)$.  We can recover the result of  Gorce et al.~\cite{Gorce2016} when $\epsilon\rightarrow 0$ in Eq.~(\ref{eq:Sv78}) ]The scaling used by them~\cite{Gorce2016} is slightly different from Eq.~(\ref{eq:Sv70})].

\begin{figure}[htbp]
\begin{center}
\includegraphics[width=0.9\linewidth]{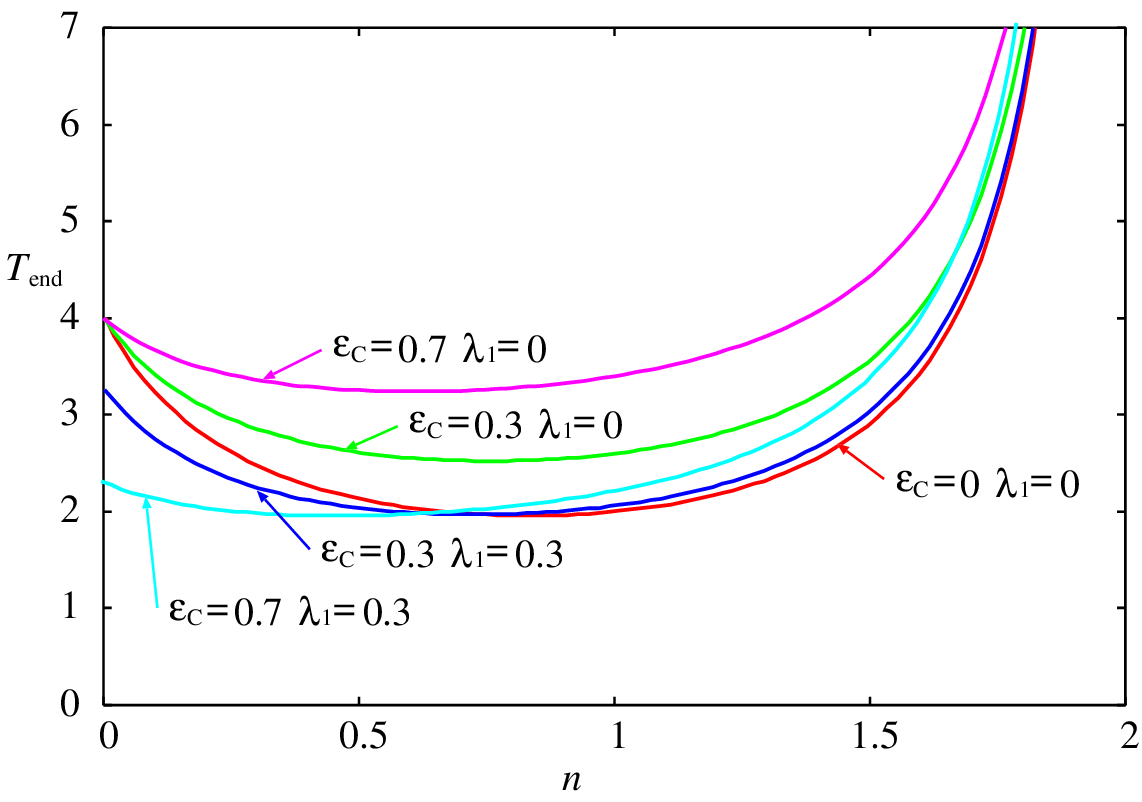}
\end{center}
\caption{
The time $T_{\rm end}$ necessary to reach the outlet $z=H (L=1)$ as a function of the exponent $n$ for various perturbation parameter $\epsilon_{\rm C}$ and the slip length coefficient $\lambda_{1}$.  The result obtained from Eq.~(\ref{eq:Sv1}) using the slowly-varying approximation corresponds to $\epsilon_{\rm C}=0$ and $\lambda_{1}=0$.  All curves show minimum so that optimal $n_{\rm min}$ exist.
}
\label{fig:SL5}
\end{figure}

Figure~\ref{fig:SL5} shows the time $T_{\rm end}$ as a function of the exponent $n$.  It has a limiting value
\begin{equation}
T_{\rm end}\rightarrow 4-8\epsilon_{\rm C}\lambda_{1},\;\;\;n\rightarrow 0,
\label{eq:Sv79}
\end{equation}
and shows a minimum, which occurs at
\begin{equation}
n_{\rm min}=\frac{-3+2\epsilon_{\rm C}\left(1+\lambda_{1}\right)+\sqrt{9-7\epsilon_{\rm C}\left(1+\lambda_{1}\right)}}{\epsilon_{\rm C}\left(1+\lambda_{1}\right)},
\label{eq:Sv80}
\end{equation}
obtained by minimizing Eq.~(\ref{eq:Sv78}).  In the slowly-varying limit
\begin{equation}
n_{\rm min}\sim\frac{5}{6}-\frac{49}{216}\epsilon_{\rm C}\left(1+\lambda_{1}\right)+\dots \rightarrow \frac{5}{6}\simeq 0.833,\;\;\;\epsilon_{\rm C}\rightarrow 0,
\label{eq:Sv81}
\end{equation}
which was derived~\cite{Gorce2016} using the slowly-varying approximation in Eq.~(\ref{eq:Sv1}).

\begin{figure}[htbp]
\begin{center}
\includegraphics[width=0.9\linewidth]{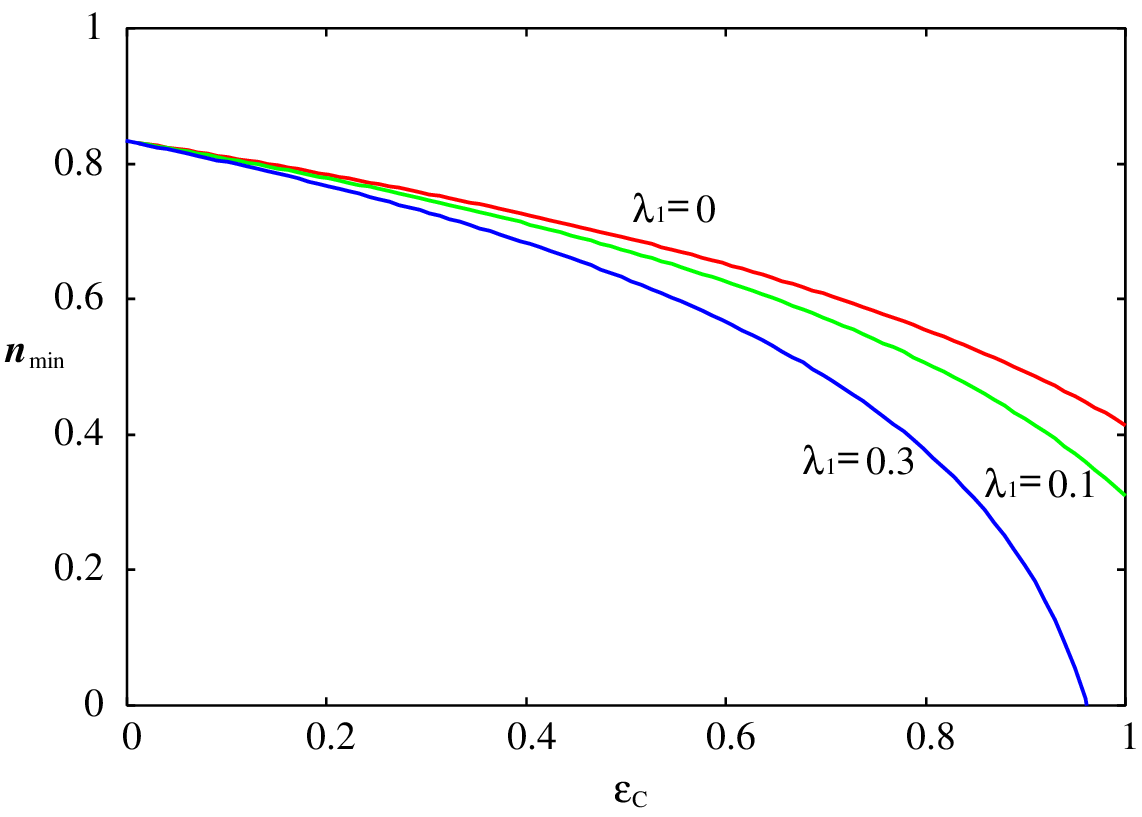}
\end{center}
\caption{
The optimum $n_{\rm min}$ at which the time $T_{\rm end}$ (Fig.~\ref{fig:SL5}) takes minimum time as a function of the non-slowly-varying perturbation parameter $\epsilon_{\rm C}$.
}
\label{fig:SL6}
\end{figure}

In Fig.~\ref{fig:SL6}, we show $n_{\rm min}$ as a function of the perturbation parameter $\epsilon_{\rm C}$.  As the non-slowly-varying perturbation is larger, the optimal exponent $n_{\rm min}$ becomes smaller so that the optimal shape of the tube is fatter [Fig.~\ref{fig:SL4}(b).  Of course, the practical problem of fabrication and fine tuning of the exponent $n$ may not be easy.

\section{\label{sec:sec4} Conclusion}

In this study, we considered the capillary flow in an axially symmetric tube of a {\it non-slowly-varying} radius. We also considered the Navier’s boundary slip and used the hydrodynamic model~\cite{Duarte1996,Digilov2008,Liou2009,Wang2013,Nissan2016} to derived macroscopic equation for the capillary flow from the Stokes equation. The new formula derived is a direct extension of the popular Hagen-Poiseulle law for the capillary flow in a tube of a {\it slowly-varying radius}~\cite{Batchelor1967} which has been widely used ~\cite{Sharma1991,Staples2002,Young2004,Reyssat2008,Urteaga2013,Berli2014,Elizalde2014,Gorce2016}.

Using this newly derived formula, we reconsidered the dynamics of meniscus by the capillary imbibition and the extension of the classical Lucas-Washburn scaling law~\cite{Lucas1918,Washburn1921} in several typical geometries.  We found that the capillary imbibition in a converging and a diverging conical tube is not affected by the non-slowly-varying effect, so that the conical tube shows a diode-like asymmetric transport~\cite{Urteaga2013,Berli2014,Singh2020,Iwamatsu2022} even if its radius varies not slowly.  We also considered the capillary imbibition in a diverging~\cite{Reyssat2008} and a converging~\cite{Gorce2016} power-law-shaped tube and identified the perturbation parameter which characterizes the non-slowly-varying radius.  In those cases, the imbibition is affected by a non-slowly-varying radius so that their scaling laws and optimal shapes could be modified if the non-slowly-varying effect is important.  Our results will supplement the popular formula~\cite{Batchelor1967} widely used and will be useful to assess the fluid transport ability of axially symmetric tubes with non-slowly-varying cross sections.

\section*{Author Declaration}
\subsection*{Conflict of interest}
The author declares no conflict of interest.

\section*{Data Availability Statement}
The data that support the findings of this study are available from the author upon reasonable request.

\bibliography{POF24V4}

\begin{thebibliography}{55}%
\makeatletter
\providecommand \@ifxundefined [1]{%
 \@ifx{#1\undefined}
}%
\providecommand \@ifnum [1]{%
 \ifnum #1\expandafter \@firstoftwo
 \else \expandafter \@secondoftwo
 \fi
}%
\providecommand \@ifx [1]{%
 \ifx #1\expandafter \@firstoftwo
 \else \expandafter \@secondoftwo
 \fi
}%
\providecommand \natexlab [1]{#1}%
\providecommand \enquote  [1]{``#1''}%
\providecommand \bibnamefont  [1]{#1}%
\providecommand \bibfnamefont [1]{#1}%
\providecommand \citenamefont [1]{#1}%
\providecommand \href@noop [0]{\@secondoftwo}%
\providecommand \href [0]{\begingroup \@sanitize@url \@href}%
\providecommand \@href[1]{\@@startlink{#1}\@@href}%
\providecommand \@@href[1]{\endgroup#1\@@endlink}%
\providecommand \@sanitize@url [0]{\catcode `\\12\catcode `\$12\catcode
  `\&12\catcode `\#12\catcode `\^12\catcode `\_12\catcode `\%12\relax}%
\providecommand \@@startlink[1]{}%
\providecommand \@@endlink[0]{}%
\providecommand \url  [0]{\begingroup\@sanitize@url \@url }%
\providecommand \@url [1]{\endgroup\@href {#1}{\urlprefix }}%
\providecommand \urlprefix  [0]{URL }%
\providecommand \Eprint [0]{\href }%
\providecommand \doibase [0]{http://dx.doi.org/}%
\providecommand \selectlanguage [0]{\@gobble}%
\providecommand \bibinfo  [0]{\@secondoftwo}%
\providecommand \bibfield  [0]{\@secondoftwo}%
\providecommand \translation [1]{[#1]}%
\providecommand \BibitemOpen [0]{}%
\providecommand \bibitemStop [0]{}%
\providecommand \bibitemNoStop [0]{.\EOS\space}%
\providecommand \EOS [0]{\spacefactor3000\relax}%
\providecommand \BibitemShut  [1]{\csname bibitem#1\endcsname}%
\let\auto@bib@innerbib\@empty
\bibitem [{\citenamefont {G.~K.~Batchelor}(1967)}]{Batchelor1967}%
  \BibitemOpen
  \bibfield  {author} {\bibinfo {author} {\bibfnamefont {F.}~\bibnamefont
  {G.~K.~Batchelor}},\ }\href@noop {} {\emph {\bibinfo {title} {An Introduction
  to Fluid Dynamics}}}\ (\bibinfo  {publisher} {Cambridge UP},\ \bibinfo
  {address} {Cambridge},\ \bibinfo {year} {1967})\BibitemShut {NoStop}%
\bibitem [{\citenamefont {Cai}\ \emph {et~al.}(2022)\citenamefont {Cai},
  \citenamefont {Chen}, \citenamefont {Liu}, \citenamefont {Li},\ and\
  \citenamefont {Sun}}]{Cai2022}%
  \BibitemOpen
  \bibfield  {author} {\bibinfo {author} {\bibfnamefont {J.}~\bibnamefont
  {Cai}}, \bibinfo {author} {\bibfnamefont {Y.}~\bibnamefont {Chen}}, \bibinfo
  {author} {\bibfnamefont {Y.}~\bibnamefont {Liu}}, \bibinfo {author}
  {\bibfnamefont {S.}~\bibnamefont {Li}}, \ and\ \bibinfo {author}
  {\bibfnamefont {C.}~\bibnamefont {Sun}},\ }\bibfield  {title} {\enquote
  {\bibinfo {title} {Capillary imbibition and flow of wetting liquid in
  irregular capillaries: A 100-year review},}\ }\href {\doibase xxx} {\bibfield
   {journal} {\bibinfo  {journal} {Adv. Colloid Interface Sci.}\ }\textbf
  {\bibinfo {volume} {304}},\ \bibinfo {pages} {102654} (\bibinfo {year}
  {2022})}\BibitemShut {NoStop}%
\bibitem [{\citenamefont {Prakash}, \citenamefont {Qu\'er\'e},\ and\
  \citenamefont {Bush}(2008)}]{Prakash2008}%
  \BibitemOpen
  \bibfield  {author} {\bibinfo {author} {\bibfnamefont {M.}~\bibnamefont
  {Prakash}}, \bibinfo {author} {\bibfnamefont {D.}~\bibnamefont {Qu\'er\'e}},
  \ and\ \bibinfo {author} {\bibfnamefont {J.~W.~M.}\ \bibnamefont {Bush}},\
  }\bibfield  {title} {\enquote {\bibinfo {title} {Surface tension transport of
  prey by feeding shorebirds: The capillary ratchet},}\ }\href {\doibase xxx}
  {\bibfield  {journal} {\bibinfo  {journal} {Science}\ }\textbf {\bibinfo
  {volume} {320}},\ \bibinfo {pages} {931} (\bibinfo {year}
  {2008})}\BibitemShut {NoStop}%
\bibitem [{\citenamefont {Kim}\ and\ \citenamefont {Bush}(2012)}]{Kim2012}%
  \BibitemOpen
  \bibfield  {author} {\bibinfo {author} {\bibfnamefont {W.}~\bibnamefont
  {Kim}}\ and\ \bibinfo {author} {\bibfnamefont {J.~W.~M.}\ \bibnamefont
  {Bush}},\ }\bibfield  {title} {\enquote {\bibinfo {title} {Natural drinking
  strategies},}\ }\href {\doibase xxx} {\bibfield  {journal} {\bibinfo
  {journal} {J. Fluid. Mech.}\ }\textbf {\bibinfo {volume} {705}},\ \bibinfo
  {pages} {7} (\bibinfo {year} {2012})}\BibitemShut {NoStop}%
\bibitem [{\citenamefont {Squires}\ and\ \citenamefont
  {Quake}(2005)}]{Squires2005}%
  \BibitemOpen
  \bibfield  {author} {\bibinfo {author} {\bibfnamefont {T.~M.}\ \bibnamefont
  {Squires}}\ and\ \bibinfo {author} {\bibfnamefont {S.~R.}\ \bibnamefont
  {Quake}},\ }\bibfield  {title} {\enquote {\bibinfo {title} {Microfluidics:
  Fluid physics at the nanoliter scale},}\ }\href {\doibase xxx} {\bibfield
  {journal} {\bibinfo  {journal} {Rev. Mod. Phys.}\ }\textbf {\bibinfo {volume}
  {77}},\ \bibinfo {pages} {977} (\bibinfo {year} {2005})}\BibitemShut
  {NoStop}%
\bibitem [{\citenamefont {Bocquet}\ and\ \citenamefont
  {Charlaix}(2010)}]{Bocquet2010}%
  \BibitemOpen
  \bibfield  {author} {\bibinfo {author} {\bibfnamefont {L.}~\bibnamefont
  {Bocquet}}\ and\ \bibinfo {author} {\bibfnamefont {E.}~\bibnamefont
  {Charlaix}},\ }\bibfield  {title} {\enquote {\bibinfo {title} {Nanofluids,
  from bulk to interfaces},}\ }\href {\doibase xxx} {\bibfield  {journal}
  {\bibinfo  {journal} {Chem. Soc. Rev.}\ }\textbf {\bibinfo {volume} {39}},\
  \bibinfo {pages} {1073} (\bibinfo {year} {2010})}\BibitemShut {NoStop}%
\bibitem [{\citenamefont {Kavokine}, \citenamefont {Netz},\ and\ \citenamefont
  {Bocquet}(2021)}]{Kavokine2021}%
  \BibitemOpen
  \bibfield  {author} {\bibinfo {author} {\bibfnamefont {N.}~\bibnamefont
  {Kavokine}}, \bibinfo {author} {\bibfnamefont {R.~R.}\ \bibnamefont {Netz}},
  \ and\ \bibinfo {author} {\bibfnamefont {L.}~\bibnamefont {Bocquet}},\
  }\bibfield  {title} {\enquote {\bibinfo {title} {Fluids at the nanoscale:
  from continuum to subcontinuum transport},}\ }\href {\doibase xxx} {\bibfield
   {journal} {\bibinfo  {journal} {Annu. Rev. Fluid Mech.}\ }\textbf {\bibinfo
  {volume} {53}},\ \bibinfo {pages} {377} (\bibinfo {year} {2021})}\BibitemShut
  {NoStop}%
\bibitem [{\citenamefont {Comanns}\ \emph {et~al.}(2015)\citenamefont
  {Comanns}, \citenamefont {Buchberger}, \citenamefont {Buchsbaum},
  \citenamefont {Baumgartner}, \citenamefont {Kogler}, \citenamefont {Bauer},\
  and\ \citenamefont {Baumgartner}}]{Comanns2015}%
  \BibitemOpen
  \bibfield  {author} {\bibinfo {author} {\bibfnamefont {P.}~\bibnamefont
  {Comanns}}, \bibinfo {author} {\bibfnamefont {G.}~\bibnamefont {Buchberger}},
  \bibinfo {author} {\bibfnamefont {A.}~\bibnamefont {Buchsbaum}}, \bibinfo
  {author} {\bibfnamefont {R.}~\bibnamefont {Baumgartner}}, \bibinfo {author}
  {\bibfnamefont {A.}~\bibnamefont {Kogler}}, \bibinfo {author} {\bibfnamefont
  {S.}~\bibnamefont {Bauer}}, \ and\ \bibinfo {author} {\bibfnamefont
  {W.}~\bibnamefont {Baumgartner}},\ }\bibfield  {title} {\enquote {\bibinfo
  {title} {Directional, passive liquid transport: the texas horned lizard as a
  model for a biomimetic 'liquid diode'},}\ }\href {\doibase xxx} {\bibfield
  {journal} {\bibinfo  {journal} {J. Roy. Soc. Interface}\ }\textbf {\bibinfo
  {volume} {12}},\ \bibinfo {pages} {20150415} (\bibinfo {year}
  {2015})}\BibitemShut {NoStop}%
\bibitem [{\citenamefont {Bell}\ and\ \citenamefont
  {Cameron}(1906)}]{Bell1906}%
  \BibitemOpen
  \bibfield  {author} {\bibinfo {author} {\bibfnamefont {J.~M.}\ \bibnamefont
  {Bell}}\ and\ \bibinfo {author} {\bibfnamefont {F.~K.}\ \bibnamefont
  {Cameron}},\ }\bibfield  {title} {\enquote {\bibinfo {title} {The flow of
  liquid through capillary spaces},}\ }\href {\doibase xxx} {\bibfield
  {journal} {\bibinfo  {journal} {J. Phys. Chem.}\ }\textbf {\bibinfo {volume}
  {10}},\ \bibinfo {pages} {658} (\bibinfo {year} {1906})}\BibitemShut
  {NoStop}%
\bibitem [{\citenamefont {Lucas}(1918)}]{Lucas1918}%
  \BibitemOpen
  \bibfield  {author} {\bibinfo {author} {\bibfnamefont {R.}~\bibnamefont
  {Lucas}},\ }\bibfield  {title} {\enquote {\bibinfo {title} {Ueber das
  zeitgesetz des kapillaren aufstiegs von fl\"ussigkeiten},}\ }\href {\doibase
  xxx} {\bibfield  {journal} {\bibinfo  {journal} {Kolloid Z.}\ }\textbf
  {\bibinfo {volume} {23}},\ \bibinfo {pages} {15} (\bibinfo {year}
  {1918})}\BibitemShut {NoStop}%
\bibitem [{\citenamefont {Washburn}(1921)}]{Washburn1921}%
  \BibitemOpen
  \bibfield  {author} {\bibinfo {author} {\bibfnamefont {E.~W.}\ \bibnamefont
  {Washburn}},\ }\bibfield  {title} {\enquote {\bibinfo {title} {The dynamics
  of capillary flow},}\ }\href {\doibase xxx} {\bibfield  {journal} {\bibinfo
  {journal} {Phys. Rev.}\ }\textbf {\bibinfo {volume} {17}},\ \bibinfo {pages}
  {273} (\bibinfo {year} {1921})}\BibitemShut {NoStop}%
\bibitem [{\citenamefont {Rideal}(1922)}]{Rideal1922}%
  \BibitemOpen
  \bibfield  {author} {\bibinfo {author} {\bibfnamefont {E.~K.}\ \bibnamefont
  {Rideal}},\ }\bibfield  {title} {\enquote {\bibinfo {title} {On the flow of
  liquids under capillary pressure},}\ }\href {\doibase xxx} {\bibfield
  {journal} {\bibinfo  {journal} {Phil. Mag.}\ }\textbf {\bibinfo {volume}
  {44}},\ \bibinfo {pages} {1152} (\bibinfo {year} {1922})}\BibitemShut
  {NoStop}%
\bibitem [{\citenamefont {Bosanquet}(1923)}]{Bosanquet1923}%
  \BibitemOpen
  \bibfield  {author} {\bibinfo {author} {\bibfnamefont {C.~H.}\ \bibnamefont
  {Bosanquet}},\ }\bibfield  {title} {\enquote {\bibinfo {title} {On the flow
  of liquids into capillary tubes},}\ }\href {\doibase xxx} {\bibfield
  {journal} {\bibinfo  {journal} {Phil. Mag.}\ }\textbf {\bibinfo {volume}
  {45}},\ \bibinfo {pages} {525} (\bibinfo {year} {1923})}\BibitemShut
  {NoStop}%
\bibitem [{\citenamefont {Landau}\ and\ \citenamefont
  {Lifshitz}(1987)}]{Landau1987}%
  \BibitemOpen
  \bibfield  {author} {\bibinfo {author} {\bibfnamefont {L.~D.}\ \bibnamefont
  {Landau}}\ and\ \bibinfo {author} {\bibfnamefont {E.~M.}\ \bibnamefont
  {Lifshitz}},\ }\href@noop {} {\emph {\bibinfo {title} {Fluid Mechanics}}},\
  \bibinfo {edition} {2nd}\ ed.\ (\bibinfo  {publisher} {Elsevier},\ \bibinfo
  {address} {Amsterdam},\ \bibinfo {year} {1987})\BibitemShut {NoStop}%
\bibitem [{\citenamefont {de~Gennes}, \citenamefont {Brochard-Wyart},\ and\
  \citenamefont {Qu\'er\'e}(2004)}]{deGennes2004}%
  \BibitemOpen
  \bibfield  {author} {\bibinfo {author} {\bibfnamefont {P.-G.}\ \bibnamefont
  {de~Gennes}}, \bibinfo {author} {\bibfnamefont {F.}~\bibnamefont
  {Brochard-Wyart}}, \ and\ \bibinfo {author} {\bibfnamefont {D.}~\bibnamefont
  {Qu\'er\'e}},\ }\href@noop {} {\emph {\bibinfo {title} {Capillarity and
  Wetting Phenomena Drops, Bubbles, Pearls, Waves}}}\ (\bibinfo  {publisher}
  {Springer},\ \bibinfo {address} {New York},\ \bibinfo {year}
  {2004})\BibitemShut {NoStop}%
\bibitem [{\citenamefont {Sharma}\ and\ \citenamefont
  {Ross}(1991)}]{Sharma1991}%
  \BibitemOpen
  \bibfield  {author} {\bibinfo {author} {\bibfnamefont {R.}~\bibnamefont
  {Sharma}}\ and\ \bibinfo {author} {\bibfnamefont {D.~S.}\ \bibnamefont
  {Ross}},\ }\bibfield  {title} {\enquote {\bibinfo {title} {Kinetics of liquid
  penetration into periodically constrained capillaries},}\ }\href {\doibase
  xxx} {\bibfield  {journal} {\bibinfo  {journal} {J. Chem. Soc. Faraday
  Trans.}\ }\textbf {\bibinfo {volume} {87}},\ \bibinfo {pages} {619} (\bibinfo
  {year} {1991})}\BibitemShut {NoStop}%
\bibitem [{\citenamefont {Staples}\ and\ \citenamefont
  {Shaffe}(2002)}]{Staples2002}%
  \BibitemOpen
  \bibfield  {author} {\bibinfo {author} {\bibfnamefont {T.~L.}\ \bibnamefont
  {Staples}}\ and\ \bibinfo {author} {\bibfnamefont {D.~G.}\ \bibnamefont
  {Shaffe}},\ }\bibfield  {title} {\enquote {\bibinfo {title} {Wicking flow in
  irregular capillaries},}\ }\href {\doibase xxx} {\bibfield  {journal}
  {\bibinfo  {journal} {Colloid Surf A}\ }\textbf {\bibinfo {volume} {204}},\
  \bibinfo {pages} {239} (\bibinfo {year} {2002})}\BibitemShut {NoStop}%
\bibitem [{\citenamefont {Young}(2004)}]{Young2004}%
  \BibitemOpen
  \bibfield  {author} {\bibinfo {author} {\bibfnamefont {W.-B.}\ \bibnamefont
  {Young}},\ }\bibfield  {title} {\enquote {\bibinfo {title} {Analysis of
  capillary flow in non-uniform cross-sectional capillaries},}\ }\href
  {\doibase xxx} {\bibfield  {journal} {\bibinfo  {journal} {Colloid Surf A}\
  }\textbf {\bibinfo {volume} {234}},\ \bibinfo {pages} {123} (\bibinfo {year}
  {2004})}\BibitemShut {NoStop}%
\bibitem [{\citenamefont {Reyssat}\ \emph {et~al.}(2008)\citenamefont
  {Reyssat}, \citenamefont {Courbin}, \citenamefont {Reyssat},\ and\
  \citenamefont {Stone}}]{Reyssat2008}%
  \BibitemOpen
  \bibfield  {author} {\bibinfo {author} {\bibfnamefont {M.}~\bibnamefont
  {Reyssat}}, \bibinfo {author} {\bibfnamefont {L.}~\bibnamefont {Courbin}},
  \bibinfo {author} {\bibfnamefont {E.}~\bibnamefont {Reyssat}}, \ and\
  \bibinfo {author} {\bibfnamefont {H.~A.}\ \bibnamefont {Stone}},\ }\bibfield
  {title} {\enquote {\bibinfo {title} {Imbibition in geometries with axial
  variation},}\ }\href {\doibase xxx} {\bibfield  {journal} {\bibinfo
  {journal} {J. Fluid Mech.}\ }\textbf {\bibinfo {volume} {615}},\ \bibinfo
  {pages} {335} (\bibinfo {year} {2008})}\BibitemShut {NoStop}%
\bibitem [{\citenamefont {Urteaga}\ \emph {et~al.}(2013)\citenamefont
  {Urteaga}, \citenamefont {Acquaroli}, \citenamefont {Koropecki},
  \citenamefont {Santos}, \citenamefont {Alba}, \citenamefont {Pallar\`es},
  \citenamefont {Marsal},\ and\ \citenamefont {Berli}}]{Urteaga2013}%
  \BibitemOpen
  \bibfield  {author} {\bibinfo {author} {\bibfnamefont {R.}~\bibnamefont
  {Urteaga}}, \bibinfo {author} {\bibfnamefont {L.~N.}\ \bibnamefont
  {Acquaroli}}, \bibinfo {author} {\bibfnamefont {R.~R.}\ \bibnamefont
  {Koropecki}}, \bibinfo {author} {\bibfnamefont {A.}~\bibnamefont {Santos}},
  \bibinfo {author} {\bibfnamefont {M.}~\bibnamefont {Alba}}, \bibinfo {author}
  {\bibfnamefont {J.}~\bibnamefont {Pallar\`es}}, \bibinfo {author}
  {\bibfnamefont {L.~F.}\ \bibnamefont {Marsal}}, \ and\ \bibinfo {author}
  {\bibfnamefont {C.~L.~A.}\ \bibnamefont {Berli}},\ }\bibfield  {title}
  {\enquote {\bibinfo {title} {Optofluidic characterization of nanoporous
  membranes},}\ }\href {\doibase xxx} {\bibfield  {journal} {\bibinfo
  {journal} {Langmuir}\ }\textbf {\bibinfo {volume} {29}},\ \bibinfo {pages}
  {2784} (\bibinfo {year} {2013})}\BibitemShut {NoStop}%
\bibitem [{\citenamefont {Berli}\ and\ \citenamefont
  {Urteaga}(2014)}]{Berli2014}%
  \BibitemOpen
  \bibfield  {author} {\bibinfo {author} {\bibfnamefont {C.~L.~A.}\
  \bibnamefont {Berli}}\ and\ \bibinfo {author} {\bibfnamefont
  {R.}~\bibnamefont {Urteaga}},\ }\bibfield  {title} {\enquote {\bibinfo
  {title} {Asymmetric capillary filling of non-newtonian power law fluids},}\
  }\href {\doibase xxx} {\bibfield  {journal} {\bibinfo  {journal} {Microfluid
  Nanofluid}\ }\textbf {\bibinfo {volume} {17}},\ \bibinfo {pages} {1079}
  (\bibinfo {year} {2014})}\BibitemShut {NoStop}%
\bibitem [{\citenamefont {Elizalde}\ \emph {et~al.}(2014)\citenamefont
  {Elizalde}, \citenamefont {Urteaga}, \citenamefont {Koropecki},\ and\
  \citenamefont {Berli}}]{Elizalde2014}%
  \BibitemOpen
  \bibfield  {author} {\bibinfo {author} {\bibfnamefont {E.}~\bibnamefont
  {Elizalde}}, \bibinfo {author} {\bibfnamefont {R.}~\bibnamefont {Urteaga}},
  \bibinfo {author} {\bibfnamefont {R.~R.}\ \bibnamefont {Koropecki}}, \ and\
  \bibinfo {author} {\bibfnamefont {C.~L.~A.}\ \bibnamefont {Berli}},\
  }\bibfield  {title} {\enquote {\bibinfo {title} {Inverse problem of capillary
  filling},}\ }\href {\doibase xxx} {\bibfield  {journal} {\bibinfo  {journal}
  {Phys. Rev. Lett.}\ }\textbf {\bibinfo {volume} {112}},\ \bibinfo {pages}
  {134502} (\bibinfo {year} {2014})}\BibitemShut {NoStop}%
\bibitem [{\citenamefont {Gorce}, \citenamefont {Hewitt},\ and\ \citenamefont
  {Vella}(2016)}]{Gorce2016}%
  \BibitemOpen
  \bibfield  {author} {\bibinfo {author} {\bibfnamefont {J.-B.}\ \bibnamefont
  {Gorce}}, \bibinfo {author} {\bibfnamefont {I.~J.}\ \bibnamefont {Hewitt}}, \
  and\ \bibinfo {author} {\bibfnamefont {D.}~\bibnamefont {Vella}},\ }\bibfield
   {title} {\enquote {\bibinfo {title} {Capillary imbibition into converging
  tubes: Beating washburn's law and the optimal imbibition of liquids},}\
  }\href {\doibase xxx} {\bibfield  {journal} {\bibinfo  {journal} {Langmuir}\
  }\textbf {\bibinfo {volume} {32}},\ \bibinfo {pages} {1560} (\bibinfo {year}
  {2016})}\BibitemShut {NoStop}%
\bibitem [{\citenamefont {Khatoon}, \citenamefont {Phirani},\ and\
  \citenamefont {Bahga}(2020)}]{Khatoon2020}%
  \BibitemOpen
  \bibfield  {author} {\bibinfo {author} {\bibfnamefont {S.}~\bibnamefont
  {Khatoon}}, \bibinfo {author} {\bibfnamefont {J.}~\bibnamefont {Phirani}}, \
  and\ \bibinfo {author} {\bibfnamefont {S.~S.}\ \bibnamefont {Bahga}},\
  }\bibfield  {title} {\enquote {\bibinfo {title} {An analytical solution of
  the inverse problem of capillary imbibition},}\ }\href {\doibase xxx}
  {\bibfield  {journal} {\bibinfo  {journal} {Phys Fluids}\ }\textbf {\bibinfo
  {volume} {32}},\ \bibinfo {pages} {041704} (\bibinfo {year}
  {2020})}\BibitemShut {NoStop}%
\bibitem [{\citenamefont {Tran-Duc}, \citenamefont {Phan-Thien},\ and\
  \citenamefont {Thamwattana}(2020)}]{Tran-Duc2020}%
  \BibitemOpen
  \bibfield  {author} {\bibinfo {author} {\bibfnamefont {T.}~\bibnamefont
  {Tran-Duc}}, \bibinfo {author} {\bibfnamefont {N.}~\bibnamefont
  {Phan-Thien}}, \ and\ \bibinfo {author} {\bibfnamefont {N.}~\bibnamefont
  {Thamwattana}},\ }\bibfield  {title} {\enquote {\bibinfo {title} {On
  permeability of corrugated pore membranes},}\ }\href {\doibase xxx}
  {\bibfield  {journal} {\bibinfo  {journal} {AIP Adv.}\ }\textbf {\bibinfo
  {volume} {10}},\ \bibinfo {pages} {045317} (\bibinfo {year}
  {2020})}\BibitemShut {NoStop}%
\bibitem [{\citenamefont {Iwamatsu}(2022)}]{Iwamatsu2022}%
  \BibitemOpen
  \bibfield  {author} {\bibinfo {author} {\bibfnamefont {M.}~\bibnamefont
  {Iwamatsu}},\ }\bibfield  {title} {\enquote {\bibinfo {title} {Thermodynamics
  and hydrodynamics of spontaneous and forced imbibition in conical
  capillaries: A theoretical study of conical liquid diode},}\ }\href {\doibase
  xxx} {\bibfield  {journal} {\bibinfo  {journal} {Phys. Fluids}\ }\textbf
  {\bibinfo {volume} {34}},\ \bibinfo {pages} {047119} (\bibinfo {year}
  {2022})}\BibitemShut {NoStop}%
\bibitem [{\citenamefont {Tran-Duc}, \citenamefont {Phan-Thien},\ and\
  \citenamefont {Wang}(2019)}]{Tran-Duc2019}%
  \BibitemOpen
  \bibfield  {author} {\bibinfo {author} {\bibfnamefont {T.}~\bibnamefont
  {Tran-Duc}}, \bibinfo {author} {\bibfnamefont {N.}~\bibnamefont
  {Phan-Thien}}, \ and\ \bibinfo {author} {\bibfnamefont {J.}~\bibnamefont
  {Wang}},\ }\bibfield  {title} {\enquote {\bibinfo {title} {A theoretical
  study of permeability enhancement for ultrafiltration ceramic membranes with
  conical pores and slippag},}\ }\href {\doibase xxx} {\bibfield  {journal}
  {\bibinfo  {journal} {Phys. Fluids}\ }\textbf {\bibinfo {volume} {31}},\
  \bibinfo {pages} {022003} (\bibinfo {year} {2019})}\BibitemShut {NoStop}%
\bibitem [{\citenamefont {Mondal}\ \emph {et~al.}(2021)\citenamefont {Mondal},
  \citenamefont {Chaudhuri}, \citenamefont {Bakli},\ and\ \citenamefont
  {Chakraborty}}]{Mondal2021}%
  \BibitemOpen
  \bibfield  {author} {\bibinfo {author} {\bibfnamefont {N.}~\bibnamefont
  {Mondal}}, \bibinfo {author} {\bibfnamefont {A.}~\bibnamefont {Chaudhuri}},
  \bibinfo {author} {\bibfnamefont {C.}~\bibnamefont {Bakli}}, \ and\ \bibinfo
  {author} {\bibfnamefont {S.}~\bibnamefont {Chakraborty}},\ }\bibfield
  {title} {\enquote {\bibinfo {title} {Upstream events dictate interfacial slip
  in geometrically converging nanopores},}\ }\href {\doibase xxx} {\bibfield
  {journal} {\bibinfo  {journal} {J. Chem. Phys.}\ }\textbf {\bibinfo {volume}
  {154}},\ \bibinfo {pages} {164709} (\bibinfo {year} {2021})}\BibitemShut
  {NoStop}%
\bibitem [{\citenamefont {Duarte}, \citenamefont {Strier},\ and\ \citenamefont
  {Zanette}(1996)}]{Duarte1996}%
  \BibitemOpen
  \bibfield  {author} {\bibinfo {author} {\bibfnamefont {A.~A.}\ \bibnamefont
  {Duarte}}, \bibinfo {author} {\bibfnamefont {D.~E.}\ \bibnamefont {Strier}},
  \ and\ \bibinfo {author} {\bibfnamefont {D.~H.}\ \bibnamefont {Zanette}},\
  }\bibfield  {title} {\enquote {\bibinfo {title} {The rise of a liquid in a
  capillary tube revisited: A hydrodynamical approach},}\ }\href {\doibase xxx}
  {\bibfield  {journal} {\bibinfo  {journal} {Am. J. Phys.}\ }\textbf {\bibinfo
  {volume} {64}},\ \bibinfo {pages} {413} (\bibinfo {year} {1996})}\BibitemShut
  {NoStop}%
\bibitem [{\citenamefont {Digilov}(2008)}]{Digilov2008}%
  \BibitemOpen
  \bibfield  {author} {\bibinfo {author} {\bibfnamefont {R.~M.}\ \bibnamefont
  {Digilov}},\ }\bibfield  {title} {\enquote {\bibinfo {title} {Capillary rise
  of a non-newtonian power law liquid: Impact of the fluid rheology and dynamic
  contact angle},}\ }\href {\doibase xxx} {\bibfield  {journal} {\bibinfo
  {journal} {Langmuir}\ }\textbf {\bibinfo {volume} {24}},\ \bibinfo {pages}
  {13663} (\bibinfo {year} {2008})}\BibitemShut {NoStop}%
\bibitem [{\citenamefont {Liou}, \citenamefont {Peng},\ and\ \citenamefont
  {Parker}(2009)}]{Liou2009}%
  \BibitemOpen
  \bibfield  {author} {\bibinfo {author} {\bibfnamefont {W.~W.}\ \bibnamefont
  {Liou}}, \bibinfo {author} {\bibfnamefont {Y.}~\bibnamefont {Peng}}, \ and\
  \bibinfo {author} {\bibfnamefont {P.~E.}\ \bibnamefont {Parker}},\ }\bibfield
   {title} {\enquote {\bibinfo {title} {Analytical modeling of capillary flow
  in tubes of nonuniform cross section},}\ }\href {\doibase xxx} {\bibfield
  {journal} {\bibinfo  {journal} {J. Colloid Interface Sci.}\ }\textbf
  {\bibinfo {volume} {333}},\ \bibinfo {pages} {389} (\bibinfo {year}
  {2009})}\BibitemShut {NoStop}%
\bibitem [{\citenamefont {Wang}, \citenamefont {Graber},\ and\ \citenamefont
  {Wallach}(2013)}]{Wang2013}%
  \BibitemOpen
  \bibfield  {author} {\bibinfo {author} {\bibfnamefont {Q.}~\bibnamefont
  {Wang}}, \bibinfo {author} {\bibfnamefont {E.~R.}\ \bibnamefont {Graber}}, \
  and\ \bibinfo {author} {\bibfnamefont {R.}~\bibnamefont {Wallach}},\
  }\bibfield  {title} {\enquote {\bibinfo {title} {Synergistic effects of
  geometry, inertia, and dynamic contact angle on wetting and dewetting of
  capillaries of varying cross sections},}\ }\href {\doibase xxx} {\bibfield
  {journal} {\bibinfo  {journal} {J. Colloid Interface Sci.}\ }\textbf
  {\bibinfo {volume} {396}},\ \bibinfo {pages} {270} (\bibinfo {year}
  {2013})}\BibitemShut {NoStop}%
\bibitem [{\citenamefont {Nissan}, \citenamefont {Wang},\ and\ \citenamefont
  {Wallach}(2016)}]{Nissan2016}%
  \BibitemOpen
  \bibfield  {author} {\bibinfo {author} {\bibfnamefont {A.}~\bibnamefont
  {Nissan}}, \bibinfo {author} {\bibfnamefont {Q.}~\bibnamefont {Wang}}, \ and\
  \bibinfo {author} {\bibfnamefont {R.}~\bibnamefont {Wallach}},\ }\bibfield
  {title} {\enquote {\bibinfo {title} {Kinetics of gravity-driven slug flow in
  partially wettable capillaries of varying cross section},}\ }\href {\doibase
  xxx} {\bibfield  {journal} {\bibinfo  {journal} {Water Resour. Res.}\
  }\textbf {\bibinfo {volume} {52}},\ \bibinfo {pages} {8472} (\bibinfo {year}
  {2016})}\BibitemShut {NoStop}%
\bibitem [{\citenamefont {Menon}\ and\ \citenamefont
  {Agrawal}(1987)}]{Menon1987}%
  \BibitemOpen
  \bibfield  {author} {\bibinfo {author} {\bibfnamefont {V.~J.}\ \bibnamefont
  {Menon}}\ and\ \bibinfo {author} {\bibfnamefont {D.~C.}\ \bibnamefont
  {Agrawal}},\ }\bibfield  {title} {\enquote {\bibinfo {title} {Newton's law of
  motion for variable mass systems applied to capillarity},}\ }\href {\doibase
  xxx} {\bibfield  {journal} {\bibinfo  {journal} {Am. J. Phys.}\ }\textbf
  {\bibinfo {volume} {55}},\ \bibinfo {pages} {63} (\bibinfo {year}
  {1987})}\BibitemShut {NoStop}%
\bibitem [{\citenamefont {Zhmud}, \citenamefont {Tiberg},\ and\ \citenamefont
  {Hallstensson}(2000)}]{Zhmud2000}%
  \BibitemOpen
  \bibfield  {author} {\bibinfo {author} {\bibfnamefont {B.~V.}\ \bibnamefont
  {Zhmud}}, \bibinfo {author} {\bibfnamefont {F.}~\bibnamefont {Tiberg}}, \
  and\ \bibinfo {author} {\bibfnamefont {K.}~\bibnamefont {Hallstensson}},\
  }\bibfield  {title} {\enquote {\bibinfo {title} {Dynamics of capillary
  rise},}\ }\href {\doibase xxx} {\bibfield  {journal} {\bibinfo  {journal} {J.
  Colloid Interface Sci.}\ }\textbf {\bibinfo {volume} {228}},\ \bibinfo
  {pages} {263} (\bibinfo {year} {2000})}\BibitemShut {NoStop}%
\bibitem [{\citenamefont {Masoodi}, \citenamefont {Languri},\ and\
  \citenamefont {Ostadhossein}(2013)}]{Masoodi2013}%
  \BibitemOpen
  \bibfield  {author} {\bibinfo {author} {\bibfnamefont {R.}~\bibnamefont
  {Masoodi}}, \bibinfo {author} {\bibfnamefont {E.}~\bibnamefont {Languri}}, \
  and\ \bibinfo {author} {\bibfnamefont {A.}~\bibnamefont {Ostadhossein}},\
  }\bibfield  {title} {\enquote {\bibinfo {title} {Dynamics of liquid rise in a
  vertical capillary tube},}\ }\href {\doibase xxx} {\bibfield  {journal}
  {\bibinfo  {journal} {J. Colloid Interface Sci.}\ }\textbf {\bibinfo {volume}
  {389}},\ \bibinfo {pages} {268} (\bibinfo {year} {2013})}\BibitemShut
  {NoStop}%
\bibitem [{\citenamefont {Majumder}\ \emph {et~al.}(2005)\citenamefont
  {Majumder}, \citenamefont {Chopra}, \citenamefont {Andrews},\ and\
  \citenamefont {Hinds}}]{Majumder2005}%
  \BibitemOpen
  \bibfield  {author} {\bibinfo {author} {\bibfnamefont {M.}~\bibnamefont
  {Majumder}}, \bibinfo {author} {\bibfnamefont {N.}~\bibnamefont {Chopra}},
  \bibinfo {author} {\bibfnamefont {R.}~\bibnamefont {Andrews}}, \ and\
  \bibinfo {author} {\bibfnamefont {B.~J.}\ \bibnamefont {Hinds}},\ }\bibfield
  {title} {\enquote {\bibinfo {title} {Enhanced flow in carbon nanotubes},}\
  }\href {\doibase xxx} {\bibfield  {journal} {\bibinfo  {journal} {Nature}\
  }\textbf {\bibinfo {volume} {438}},\ \bibinfo {pages} {44} (\bibinfo {year}
  {2005})}\BibitemShut {NoStop}%
\bibitem [{\citenamefont {Neto}\ \emph {et~al.}(2005)\citenamefont {Neto},
  \citenamefont {Evans}, \citenamefont {Bonaccurso}, \citenamefont {Butt},\
  and\ \citenamefont {Craig}}]{Neto2005}%
  \BibitemOpen
  \bibfield  {author} {\bibinfo {author} {\bibfnamefont {C.}~\bibnamefont
  {Neto}}, \bibinfo {author} {\bibfnamefont {D.~R.}\ \bibnamefont {Evans}},
  \bibinfo {author} {\bibfnamefont {E.}~\bibnamefont {Bonaccurso}}, \bibinfo
  {author} {\bibfnamefont {H.-J.}\ \bibnamefont {Butt}}, \ and\ \bibinfo
  {author} {\bibfnamefont {V.~S.~J.}\ \bibnamefont {Craig}},\ }\bibfield
  {title} {\enquote {\bibinfo {title} {Boundary slip in newtonian liquids: a
  review of experimental studies},}\ }\href {\doibase xxx} {\bibfield
  {journal} {\bibinfo  {journal} {Rep. Prog. Phys.}\ }\textbf {\bibinfo
  {volume} {68}},\ \bibinfo {pages} {2859} (\bibinfo {year}
  {2005})}\BibitemShut {NoStop}%
\bibitem [{\citenamefont {Suk}\ and\ \citenamefont {Aluru}(2017)}]{Suk2017}%
  \BibitemOpen
  \bibfield  {author} {\bibinfo {author} {\bibfnamefont {M.~E.}\ \bibnamefont
  {Suk}}\ and\ \bibinfo {author} {\bibfnamefont {N.~R.}\ \bibnamefont
  {Aluru}},\ }\bibfield  {title} {\enquote {\bibinfo {title} {Modeling water
  flow through carbon nanotube membranes with entrance/exit effects},}\ }\href
  {\doibase xxx} {\bibfield  {journal} {\bibinfo  {journal} {Nanoscale
  Microscale Thermophys. Eng.}\ }\textbf {\bibinfo {volume} {21}},\ \bibinfo
  {pages} {247} (\bibinfo {year} {2017})}\BibitemShut {NoStop}%
\bibitem [{\citenamefont {Wu}\ \emph {et~al.}(2017)\citenamefont {Wu},
  \citenamefont {Chena}, \citenamefont {Lib}, \citenamefont {Lib},
  \citenamefont {Xua},\ and\ \citenamefont {Dong}}]{Wu2017c}%
  \BibitemOpen
  \bibfield  {author} {\bibinfo {author} {\bibfnamefont {K.}~\bibnamefont
  {Wu}}, \bibinfo {author} {\bibfnamefont {Z.}~\bibnamefont {Chena}}, \bibinfo
  {author} {\bibfnamefont {J.}~\bibnamefont {Lib}}, \bibinfo {author}
  {\bibfnamefont {X.}~\bibnamefont {Lib}}, \bibinfo {author} {\bibfnamefont
  {J.}~\bibnamefont {Xua}}, \ and\ \bibinfo {author} {\bibfnamefont
  {X.}~\bibnamefont {Dong}},\ }\bibfield  {title} {\enquote {\bibinfo {title}
  {Intrusion and extrusion of water in hydrophobic nanopores},}\ }\href
  {\doibase xxx} {\bibfield  {journal} {\bibinfo  {journal} {Proc. Natl. Acad.
  Sci. USA}\ }\textbf {\bibinfo {volume} {114}},\ \bibinfo {pages} {3358}
  (\bibinfo {year} {2017})}\BibitemShut {NoStop}%
\bibitem [{\citenamefont {Wu}, \citenamefont {Nikolov},\ and\ \citenamefont
  {Wasan}(017a)}]{Wu2017b}%
  \BibitemOpen
  \bibfield  {author} {\bibinfo {author} {\bibfnamefont {P.}~\bibnamefont
  {Wu}}, \bibinfo {author} {\bibfnamefont {A.}~\bibnamefont {Nikolov}}, \ and\
  \bibinfo {author} {\bibfnamefont {D.}~\bibnamefont {Wasan}},\ }\bibfield
  {title} {\enquote {\bibinfo {title} {Capillary dynamics driven by molecular
  self-layering},}\ }\href {\doibase xxx} {\bibfield  {journal} {\bibinfo
  {journal} {Adv. Colloid Interface Sci.}\ }\textbf {\bibinfo {volume} {243}},\
  \bibinfo {pages} {114} (\bibinfo {year} {2017a})}\BibitemShut {NoStop}%
\bibitem [{\citenamefont {Popescu}, \citenamefont {Ralston},\ and\
  \citenamefont {Sedev}(2008)}]{Popescu2008}%
  \BibitemOpen
  \bibfield  {author} {\bibinfo {author} {\bibfnamefont {M.~N.}\ \bibnamefont
  {Popescu}}, \bibinfo {author} {\bibfnamefont {J.}~\bibnamefont {Ralston}}, \
  and\ \bibinfo {author} {\bibfnamefont {R.}~\bibnamefont {Sedev}},\ }\bibfield
   {title} {\enquote {\bibinfo {title} {Capillary rise with velocity-dependent
  dynamic contact angle},}\ }\href {\doibase xxx} {\bibfield  {journal}
  {\bibinfo  {journal} {Langmuir}\ }\textbf {\bibinfo {volume} {24}},\ \bibinfo
  {pages} {12710} (\bibinfo {year} {2008})}\BibitemShut {NoStop}%
\bibitem [{\citenamefont {Wu}, \citenamefont {Nikolov},\ and\ \citenamefont
  {Wasan}(017b)}]{Wu2017}%
  \BibitemOpen
  \bibfield  {author} {\bibinfo {author} {\bibfnamefont {P.}~\bibnamefont
  {Wu}}, \bibinfo {author} {\bibfnamefont {A.~D.}\ \bibnamefont {Nikolov}}, \
  and\ \bibinfo {author} {\bibfnamefont {D.~T.}\ \bibnamefont {Wasan}},\
  }\bibfield  {title} {\enquote {\bibinfo {title} {Capillary rise: Validity of
  the dynamic contact angle models},}\ }\href {\doibase xxx} {\bibfield
  {journal} {\bibinfo  {journal} {Langmuir}\ }\textbf {\bibinfo {volume}
  {33}},\ \bibinfo {pages} {7862} (\bibinfo {year} {2017b})}\BibitemShut
  {NoStop}%
\bibitem [{\citenamefont {Sampson}(1891)}]{Sampson1891}%
  \BibitemOpen
  \bibfield  {author} {\bibinfo {author} {\bibfnamefont {R.~A.}\ \bibnamefont
  {Sampson}},\ }\bibfield  {title} {\enquote {\bibinfo {title} {On stokes's
  current function},}\ }\href@noop {} {\bibfield  {journal} {\bibinfo
  {journal} {Phil. Trans. Roy. Soc. London A}\ }\textbf {\bibinfo {volume}
  {182}},\ \bibinfo {pages} {449} (\bibinfo {year} {1891})}\BibitemShut
  {NoStop}%
\bibitem [{\citenamefont {Weissberg}(1962)}]{Weissberg1962}%
  \BibitemOpen
  \bibfield  {author} {\bibinfo {author} {\bibfnamefont {H.~L.}\ \bibnamefont
  {Weissberg}},\ }\bibfield  {title} {\enquote {\bibinfo {title} {End
  correction for slow viscous flow through long tubes},}\ }\href {\doibase xxx}
  {\bibfield  {journal} {\bibinfo  {journal} {Phys. Fluids}\ }\textbf {\bibinfo
  {volume} {5}},\ \bibinfo {pages} {1033} (\bibinfo {year} {1962})}\BibitemShut
  {NoStop}%
\bibitem [{\citenamefont {Kornev}\ and\ \citenamefont
  {Neimark}(2000)}]{Kornev2000}%
  \BibitemOpen
  \bibfield  {author} {\bibinfo {author} {\bibfnamefont {K.~G.}\ \bibnamefont
  {Kornev}}\ and\ \bibinfo {author} {\bibfnamefont {A.~V.}\ \bibnamefont
  {Neimark}},\ }\bibfield  {title} {\enquote {\bibinfo {title} {Spontaneous
  penetration of liquids into capillaries and porous membranes revisited},}\
  }\href {\doibase xxx} {\bibfield  {journal} {\bibinfo  {journal} {J. Colloid
  Interface Sci.}\ }\textbf {\bibinfo {volume} {235}},\ \bibinfo {pages} {101}
  (\bibinfo {year} {2000})}\BibitemShut {NoStop}%
\bibitem [{\citenamefont {Stange}, \citenamefont {Dreyer},\ and\ \citenamefont
  {Rath}(2003)}]{Stange2003}%
  \BibitemOpen
  \bibfield  {author} {\bibinfo {author} {\bibfnamefont {M.}~\bibnamefont
  {Stange}}, \bibinfo {author} {\bibfnamefont {M.~E.}\ \bibnamefont {Dreyer}},
  \ and\ \bibinfo {author} {\bibfnamefont {H.~J.}\ \bibnamefont {Rath}},\
  }\bibfield  {title} {\enquote {\bibinfo {title} {Capillary driven flow in
  circular cylindrical tubes},}\ }\href {\doibase xxx} {\bibfield  {journal}
  {\bibinfo  {journal} {Phys. Fluids}\ }\textbf {\bibinfo {volume} {15}},\
  \bibinfo {pages} {2587} (\bibinfo {year} {2003})}\BibitemShut {NoStop}%
\bibitem [{\citenamefont {Willmott}, \citenamefont {Neto},\ and\ \citenamefont
  {Hendy}(2011)}]{Willmott2011}%
  \BibitemOpen
  \bibfield  {author} {\bibinfo {author} {\bibfnamefont {G.~R.}\ \bibnamefont
  {Willmott}}, \bibinfo {author} {\bibfnamefont {C.}~\bibnamefont {Neto}}, \
  and\ \bibinfo {author} {\bibfnamefont {S.~C.}\ \bibnamefont {Hendy}},\
  }\bibfield  {title} {\enquote {\bibinfo {title} {Uptake of water droplets by
  nonwetting capillaries},}\ }\href {\doibase xxx} {\bibfield  {journal}
  {\bibinfo  {journal} {Soft Matters}\ }\textbf {\bibinfo {volume} {7}},\
  \bibinfo {pages} {2357} (\bibinfo {year} {2011})}\BibitemShut {NoStop}%
\bibitem [{\citenamefont {Scholle}, \citenamefont {Wierschem},\ and\
  \citenamefont {Aksel}(2004)}]{Scholle2004}%
  \BibitemOpen
  \bibfield  {author} {\bibinfo {author} {\bibfnamefont {M.}~\bibnamefont
  {Scholle}}, \bibinfo {author} {\bibfnamefont {A.}~\bibnamefont {Wierschem}},
  \ and\ \bibinfo {author} {\bibfnamefont {N.}~\bibnamefont {Aksel}},\
  }\bibfield  {title} {\enquote {\bibinfo {title} {Creeping films with vortices
  over strongly undulated bottoms},}\ }\href {\doibase xxx} {\bibfield
  {journal} {\bibinfo  {journal} {Acta Mech.}\ }\textbf {\bibinfo {volume}
  {168}},\ \bibinfo {pages} {167} (\bibinfo {year} {2004})}\BibitemShut
  {NoStop}%
\bibitem [{\citenamefont {Marner}, \citenamefont {Gaskell},\ and\ \citenamefont
  {Scholle}(2017)}]{Marner2017}%
  \BibitemOpen
  \bibfield  {author} {\bibinfo {author} {\bibfnamefont {F.}~\bibnamefont
  {Marner}}, \bibinfo {author} {\bibfnamefont {P.~H.}\ \bibnamefont {Gaskell}},
  \ and\ \bibinfo {author} {\bibfnamefont {M.}~\bibnamefont {Scholle}},\
  }\bibfield  {title} {\enquote {\bibinfo {title} {A complex-valued first
  integral of navier-stokes equations: Unsteady couette flow in a corrugated
  channel system},}\ }\href {\doibase xxx} {\bibfield  {journal} {\bibinfo
  {journal} {J. Math. Phys.}\ }\textbf {\bibinfo {volume} {58}},\ \bibinfo
  {pages} {043102} (\bibinfo {year} {2017})}\BibitemShut {NoStop}%
\bibitem [{\citenamefont {Goli}, \citenamefont {Saha},\ and\ \citenamefont
  {Agrawal}(2022)}]{Goli2022}%
  \BibitemOpen
  \bibfield  {author} {\bibinfo {author} {\bibfnamefont {S.}~\bibnamefont
  {Goli}}, \bibinfo {author} {\bibfnamefont {S.~K.}\ \bibnamefont {Saha}}, \
  and\ \bibinfo {author} {\bibfnamefont {A.}~\bibnamefont {Agrawal}},\
  }\bibfield  {title} {\enquote {\bibinfo {title} {Physics of fluid flow in an
  hourglass (converging-diverging) microchannel},}\ }\href {\doibase xxx}
  {\bibfield  {journal} {\bibinfo  {journal} {Phys. Fluids}\ }\textbf {\bibinfo
  {volume} {34}},\ \bibinfo {pages} {052006} (\bibinfo {year}
  {2022})}\BibitemShut {NoStop}%
\bibitem [{\citenamefont {Singh}, \citenamefont {Kumar},\ and\ \citenamefont
  {Khan}(2020)}]{Singh2020}%
  \BibitemOpen
  \bibfield  {author} {\bibinfo {author} {\bibfnamefont {M.}~\bibnamefont
  {Singh}}, \bibinfo {author} {\bibfnamefont {A.}~\bibnamefont {Kumar}}, \ and\
  \bibinfo {author} {\bibfnamefont {A.~R.}\ \bibnamefont {Khan}},\ }\bibfield
  {title} {\enquote {\bibinfo {title} {Capillary as a liquid diode},}\ }\href
  {\doibase xxx} {\bibfield  {journal} {\bibinfo  {journal} {Phys. Rev.
  Fluids}\ }\textbf {\bibinfo {volume} {5}},\ \bibinfo {pages} {102101(R)}
  (\bibinfo {year} {2020})}\BibitemShut {NoStop}%
\bibitem [{\citenamefont {Panter}, \citenamefont {Gizaw},\ and\ \citenamefont
  {Kusumaatmaja}(2020)}]{Panter2020}%
  \BibitemOpen
  \bibfield  {author} {\bibinfo {author} {\bibfnamefont {J.~R.}\ \bibnamefont
  {Panter}}, \bibinfo {author} {\bibfnamefont {Y.}~\bibnamefont {Gizaw}}, \
  and\ \bibinfo {author} {\bibfnamefont {H.}~\bibnamefont {Kusumaatmaja}},\
  }\bibfield  {title} {\enquote {\bibinfo {title} {Critical pressure asymmetry
  in the enclosed fluid diode},}\ }\href {\doibase xxx} {\bibfield  {journal}
  {\bibinfo  {journal} {Langmuir}\ }\textbf {\bibinfo {volume} {36}},\ \bibinfo
  {pages} {7463} (\bibinfo {year} {2020})}\BibitemShut {NoStop}%
\bibitem [{\citenamefont {Xu}\ \emph {et~al.}(2023)\citenamefont {Xu},
  \citenamefont {Min}, \citenamefont {Ding}, \citenamefont {Chen},\ and\
  \citenamefont {Zhu}}]{Xu2023}%
  \BibitemOpen
  \bibfield  {author} {\bibinfo {author} {\bibfnamefont {B.}~\bibnamefont
  {Xu}}, \bibinfo {author} {\bibfnamefont {S.}~\bibnamefont {Min}}, \bibinfo
  {author} {\bibfnamefont {Y.}~\bibnamefont {Ding}}, \bibinfo {author}
  {\bibfnamefont {H.}~\bibnamefont {Chen}}, \ and\ \bibinfo {author}
  {\bibfnamefont {Q.}~\bibnamefont {Zhu}},\ }\bibfield  {title} {\enquote
  {\bibinfo {title} {Multi-pores janus paper with unidirectional liquid
  transport property toward information encryption/decryption},}\ }\href
  {\doibase xxx} {\bibfield  {journal} {\bibinfo  {journal} {Colloid Surf. A}\
  }\textbf {\bibinfo {volume} {664}},\ \bibinfo {pages} {131133} (\bibinfo
  {year} {2023})}\BibitemShut {NoStop}%
\bibitem [{\citenamefont {Iwamatsu}(2023)}]{Iwamatsu2023}%
  \BibitemOpen
  \bibfield  {author} {\bibinfo {author} {\bibfnamefont {M.}~\bibnamefont
  {Iwamatsu}},\ }\bibfield  {title} {\enquote {\bibinfo {title} {Thermodynamics
  of imbibition in capillaries of double conical structures-hourglass, diamond,
  and sawtooth shaped capillaries},}\ }\href {\doibase xxx} {\bibfield
  {journal} {\bibinfo  {journal} {Phys. Fluids}\ }\textbf {\bibinfo {volume}
  {35}},\ \bibinfo {pages} {092009} (\bibinfo {year} {2023})}\BibitemShut
  {NoStop}%
\end{thebibliography}%

\end{document}